\documentclass[aps,nofootinbib,superscriptaddress, showpacs,preprintnumbers,  nofootinbibt,twocolumn]{revtex4-2}
    \usepackage{amsmath}
    \usepackage{color}
    \usepackage{soul}
    \usepackage{epsfig}
    \usepackage{multirow}
    \usepackage{adjustbox}
    \usepackage{eurosym}
    \usepackage{dcolumn}
    \usepackage{bm}
    \usepackage{enumerate}
    \usepackage{float}
    \usepackage{epstopdf}
    \usepackage{amsmath}
    \usepackage{bm}
    \usepackage{natbib}
    \usepackage{amsfonts}
    \usepackage{amssymb}
    \usepackage{graphicx}
    \usepackage{alphalph,mathtools}
    \usepackage{etoolbox}
    \usepackage{color}
    \usepackage{booktabs}
    \usepackage{footnote}
    \usepackage{makecell,tabularx}
    \usepackage{hyperref}
    \usepackage{subfigure, rotating, bm, array}
    \hypersetup{colorlinks,citecolor=blue}
    \hypersetup{colorlinks=true,linkcolor=magenta,filecolor=magenta,
        urlcolor=blue}

    \def\be{\begin{equation}}
    \def\ee{\end{equation}}
    \def\bea{\begin{eqnarray}}
    \def\eea{\end{eqnarray}}

\begin{document}

    \title{\bf Exploring Late-Time Cosmic Acceleration with EoS Parameterizations in Horava-Lifshitz Gravity via Baryon Acoustic Oscillations}

    \author{Madhur Khurana}
    \email{K.madhur2000@gmail.com}
    \affiliation{Department of Applied Physics, Delhi Technological University, Delhi-110042, India}
    \author{Himanshu Chaudhary}
    \email{himanshuch1729@gmail.com}
    \affiliation{Department of Applied Mathematics, Delhi Technological University, Delhi-110042, India}
    \affiliation{Pacif Institute of Cosmology and Selfology (PICS), Sagara, Sambalpur 768224, Odisha, India}
    \affiliation{Department of Mathematics, Shyamlal College, University of Delhi, Delhi-110032, India.}
    \author{Ujjal Debnath}
    \email{ujjaldebnath@gmail.com}
    \affiliation{Department of
    Mathematics, Indian Institute of Engineering Science and
    Technology, Shibpur, Howrah-711 103, India.}
    \author{Alok Sardar}
    \email{alokmath94@gmail.com}
    \affiliation{Department of
    Mathematics, Indian Institute of Engineering Science and
    Technology, Shibpur, Howrah-711 103, India.}
    \author{G.Mustafa}
    \email{gmustafa3828@gmail.com}
    \affiliation{Department of Physics,
    Zhejiang Normal University, Jinhua 321004, Peoples Republic of China,}
    \affiliation{New Uzbekistan University, Mustaqillik ave. 54, 100007 Tashkent, Uzbekistan,}

\begin{abstract}
In our study, we have adopted the framework of Horava-Lifshitz gravity to model the Universe's dark matter and dark energy components. Specifically, we have considered two recent parametrizations for dark energy models: the CBDRM-type and CADMM-type parameterizations. In our analysis, we have explicitly expressed the Hubble parameter, denoted as $H(z)$, for these two distinct dark energy models. By doing so, we have aimed to investigate and quantify the accelerated cosmic expansion rate characterizing the late-time Universe. Our study uses a wide range of datasets. This dataset consists of recent measurements of baryon acoustic oscillations (BAO) collected over a period of twenty years with the Cosmic Chronometers (CC) dataset, Type Ia supernovae (SNIa) dataset, the Hubble diagram of gamma-ray bursts (GRBs), quasars (Q), and the latest measurement of the Hubble constant (R22). Consequently, we present a crucial aspect of our study by plotting the $r_{d}$ vs. $H_{0}$ plane. In the context of the $\Lambda$CDM model, after incorporating all the datasets, including the R22 prior, we obtain the following results: $H_{0}$ = $71.674089 \pm 0.734089$ $km s^{-1} Mpc^{-1}$ and $r_d = 143.050380 Mpc \pm 3.702038$. For the CBDRM model, we find $H_{0}$ = $72.355058 \pm 1.004604$ $km s^{-1} Mpc^{-1}$ and $r_d = 144.835069 Mpc \pm 2.378848$. In the case of the CADMM model, our analysis yields $H_{0}$ = $72.347804 \pm 0.923328$ $km s^{-1} Mpc^{-1}$ and $r_d = 144.466836 Mpc \pm 4.288758$. We have conducted cosmographic analyses for both of the proposed parameterizations in comparison to the $\Lambda$CDM paradigm. Additionally, we have applied Diagnostic tests to investigate the evolution of both models. Finally, the Information Criteria test demonstrates that the $\Lambda$CDM model emerges as the preferred choice among the models we have considered.
\end{abstract}

\maketitle
\tableofcontents
\section{Introduction}\label{sec1}
In the quest to unravel the mysteries of our Universe's origins and evolution, cosmologists have precisely estimated and constrained values of cosmological parameters. The precision achieved through various observational experiments \cite{1,2} has been nothing short of remarkable. Among these parameters, the Hubble constant ($H_0$) holds a special place, the cornerstone of our understanding of the cosmos. The Planck 2018 mission, representing a pinnacle of cosmic observations, has set the highest standard for determining important cosmological parameters. However, it's crucial to understand that the Planck satellite doesn't directly measure a vital parameter called the Hubble constant ($H_0$). Instead, we arrive at this value by carefully considering various cosmic factors and conducting a thorough global analysis that relies on the $\Lambda$CDM model as its foundation. In $\Lambda$CDM model, a significant study \cite{1} determined that the Hubble constant ($H_0$) is around $67.4 \pm 0.5$ km/s/Mpc. This measurement is incredibly precise, with an uncertainty of less than 1 km/s/Mpc. It strongly supported the standard cosmological model and boosted our confidence in it. However, things take an intriguing turn as we peer closer to our cosmic neighborhood, especially at lower redshifts. Measurements of $H_0$ from various sources \cite{3,4,5,6,7} begin to deviate from this precision measurement. This divergence presents us with an opportunity, albeit ironic, to explore alternative cosmological models that go beyond the familiar $\Lambda$CDM model. Enter the SHOES project \cite{4}, which employs a distance ladder method relying on celestial beacons known as cepheid stars to estimate $H_0$. Their ongoing efforts have led to a refined value of $H_0$: $73.04 \pm 1.04$ km/s/Mpc \cite{7}, departing from the earlier precision. The difference between the exact early Universe measurement \cite{1} and the more recent local measurements \cite{7} has created a puzzle in the field of cosmology known as the ``Hubble tension". This significant tension fluctuates between 4 and 5.7 $\sigma$. It urges us to examine our measurements for possible errors or calibration problems carefully. Additionally, it opens up the intriguing possibility that we may need to reevaluate the foundation of our standard cosmological model, the $\Lambda$CDM model. In essence, the Hubble tension isn't merely a scientific puzzle; it's a clarion call for cosmologists to delve deeper into the cosmos, explore its enigmas, and be open to the possibility that our understanding of the Universe's grand tapestry might yet hold surprises that challenge our current beliefs. Moreover, this intriguing tension between the early Universe measurements \cite{1} and the more recent local measurements \cite{7} serves as a tantalizing hint that new physics might be waiting to be discovered beyond our current understanding, encapsulated in the standard model. This has sparked a wave of innovative proposals for alternative cosmological models, each aiming to reconcile the disparities observed in various data surveys \cite{8,9,10,11,12,13,14,15,16,17,18,19,20}. On the other end of the spectrum, numerous endeavors have been undertaken to gauge the value of the Hubble constant through diverse observational avenues. These include astrophysical phenomena like quasar lensing \cite{21,22}, transformative events like gravitational-wave detections \cite{23,24,25}, enigmatic cosmic signals like fast radio bursts (FRBs) \cite{26,27}, and even cosmic beacons like Megamasers \cite{28,29,30}. Additionally, techniques like the red giant branch tip method (TRGS) \cite{31,32,33} and baryon acoustic oscillations (BAOs) \cite{34} have been employed to refine our estimates.  For instance, the H0LiCOW research group has ingeniously utilized the time delay caused by gravitational lensing effects to estimate $H_{0}$. In a flat $\Lambda$CDM setting, their estimations have yielded a value of $H_{0}=73.3_{-1.8}^{+1.7} \mathrm{~km} \mathrm{~s}^{-1} \mathrm{Mpc}^{-1}$ \cite{37}. Moreover, the Advanced LIGO and Virgo research teams, following the detection of the gravitational-wave event GW170817 from the merger of a binary neutron-star system, have provided a Hubble constant estimate of $H_{0}=70_{-8.0}^{+12.0} \mathrm{~km}$ $\mathrm{s}^{-1} \mathrm{Mpc}^{-1}$ \cite{38}. These observations come with a distinct advantage: they operate independently of the cosmic microwave background (CMB) surveys and the measurements obtained from the distance ladder, offering an alternative perspective on the enigmatic $H_{0}$ tensions observed in these different methodologies. Baryon acoustic oscillations (BAOs), a critical focus of our investigation, are fascinating. They represent sound waves traversing through the early Universe's baryon-photon fluid and frozen during the epoch of recombination. These oscillations manifest as distinctive patterns in the distribution of large-scale cosmic structures, and they've been discerned through various independent observational surveys. The outcomes of these BAO surveys are typically expressed in terms of $D_{A}(z) / r_{d}$, $D_{V}(z) / r_{d}$, and $H(z) \cdot r_{d}$, where $r_{d}$ denotes the comoving size of the sound horizon at the drag epoch. To delve deeper, let's rewind to the recombination era. At this pivotal juncture, photons and baryons embarked on different trajectories. Photons decoupled from baryons first, at around $z_{*} \approx 1090$, giving rise to the cosmic microwave background (CMB) we observe today. However, baryons didn't experience the drag of photons until a slightly later time, approximately at $z_{d} \approx 1059$. This significant phase transition serves as the standard ruler for BAOs. Now, here's where it gets intriguing. The Hubble constant $H_{0}$ and the sound horizon $r_{d}$ share a profound connection, together shaping what's known as the $H_{0}-r_{d}$ plane, effectively bridging the early and late epochs of the Universe's evolution. In essence, the value of $r_{d}$ hinges on the physical conditions of the early Universe, which can be finely constrained through,> precise CMB observations. However, it's worth noting that in most BAO measurements, the determination of the Hubble constant through BAO data isn't entirely free from the influence of CMB data \cite{39}. Instead of relying solely on the early-time physical calibration of $r_{d},$ an alternative approach involves combining BAO measurements with other observations from the low-redshift Universe. This strategy offers a distinct pathway to unravel the mysteries of cosmic expansion, potentially independent of the cosmic microwave background data.\\\\
One of the most challenging issues in the cosmological model is the Big Bang singularity, which arises from the general theory of relativity (GR) \citep{mukhanov2005physical}. The Big Bang theory based on GR faces several problems, including the horizon problem, the flatness problem, and the inability to explain some observed facts in the early and late Universe. The inflation theory is a consistent theory that describes the early time of the Universe, but it also suffers from the initial singularity problem at the origin (i.e., at t=0) \citep{Linde:2014nna,Lyth:1998xn,Borde:1993xh}. To avoid these problems, many physicists have sought to modify or extend GR with quantum effects or extra dimensions, hoping to find a theory free of singularities and consistent with observations. Horava-Lifshitz Gravity is one such modification of GR based on the idea of anisotropic scaling between space and time. It is a compelling alternative to Einstein's General Theory of Relativity, offering fresh insights into the fundamental nature of gravity. Developed by physicist Petr Horava, this theory emerged as an attempt to bridge the gap between general relativity and quantum mechanics, a challenge that has long puzzled physicists. The key departure from general relativity lies in the theory's introduction of anisotropic scaling, known as "Lifshitz scaling," which breaks the symmetry between space and time. Unlike Einstein's theory, which treats all directions equally, Horava-Lifshitz's Gravity implies that not all directions in spacetime are equivalent. Another distinctive feature is the incorporation of higher spatial derivatives in its gravitational action. This property makes Horava-Lifshitz Gravity potentially renormalizable, which could be significant in the quest for a quantum theory of gravity. Beyond its theoretical appeal, Horava-Lifshitz Gravity has practical implications for our understanding of the Universe. It may provide new insights into topics like the problems of dark energy \citep{Saridakis:2009bv,park2010test,Chaichian:2010yi,jamil2010new}, the behavior of black holes \citep{Cai:2009pe,danielsson2009black}, and the challenges associated with observational constraints on the parameters of the theory \citep{dutta2009observational,Dutta:2010jh}. Despite many ambiguities regarding foundational and conceptual issues of HL gravity \citep{Bogdanos:2009uj,li2009trouble}, the different cosmological scenario has been examined by several authors \citep{Pasqua:2015bfz,Karami:2011bm,Ali:2011sv,Paul:2012zz,Khodadi:2014lua,Pourhassan:2014mfa}.\\\\
In this paper, we will explore the mystery of cosmic acceleration by using a parameterized form of the dark energy equation of state (EoS) instead of relying on specific dark energy models. This way, we can avoid some problems with assuming a particular model, such as divergence issues and the possibility of misrepresenting the true nature of dark energy. However, this method also has limitations, such as the choice of parameterization and its validity over different redshift ranges. Since we do not have a clear and satisfactory theoretical model that can explain the whole evolution of the Universe, using a parameterized form of the EoS can help us learn more about the expansion history \citep{sardar2023cosmography}. Researcher have carried out numerous investigations to understand why the Universe is expanding at an accelerating rate. They have used different methods, like Equation of State (EoS) parameterizations, to explain this phenomenon \cite{17:2005pa,18-Rivera:2019aol,19:2020rho,para5}. Building on these EoS parameterizations, researchers have also thoroughly examined the deceleration parameter in their studies \cite{cunha2008transition,delCampo:2012ya,Cunha:2008mt,nair2012cosmokinetics,Xu:2007gvk,Xu:2009zza,Santos:2010gp,Turner:2001mx,Akarsu:2013lya,para1,para2,para3,para4,para6,para7,para8,para9,para10,para11,khurana2023analyzing}.
Recently, in \citep{para6}. explored a new parameter parametrization of the deceleration parameter and estimated the best-fit values for the free parameters using the chi-square and a Markov Chain Monte Carlo (MCMC) technique. Moreover, they displayed unique behaviors in the jerk, snap, statefinder, and $O_m$ diagnostic parameters for the assumed cosmological models. In \citep{para8}, the authors studied three parametrizations of the deceleration parameter and extracted the unknown constraints using different observational data. As a consequence of the above two works, we have considered two parameterizations (CBDRM and CADMM) of the EoS parameter identical to the parametrizations of the deceleration parameter in \citep{Chaudhary:2023ddn,Chaudhary:2023vxz}.\\\\
In HL cosmology, the equation of state (EoS) parameter of modified Chaplygin gas has been explored by various observational constraints \citep{Paul:2012cxu}. The cosmological scenario of the parametrizations of EoS in Chern–Simons modified gravity has been explored by Ref. \citep{jawad2019cosmological}. Also, Biswas and Debnath in Ref. \citep{Biswas:2014ooo} studied three parameterizations (linear,  CPL, and JBP) of dark energy EoS with observational constraints in the context of Horava-Lifshitz's Gravity. Using Stern, Stern+BAO and Stern+BAO+CMB joint data analysis, they obtained the bounds of the coefficient parameters $\omega_0$ and $\omega_1$. Moreover, they obtained the best-fit values of the coefficient parameters and minimum values of $\chi^2$ utilizing observational data analysis. Getting motivated by the above work, we would like to study various cosmological models, including the standard $\Lambda$CDM model, the CBDRM model, and the CADMM model, to shed light on their respective abilities to describe the evolution of the Universe. Our analysis incorporated a diverse set of cosmological datasets, encompassing Cosmic Chronometers (CC), Type Ia Supernovae (SNIa), Quasars (Q), Gamma-Ray Bursts (GRB), and the Baryon Acoustic Oscillation (BAO) scale. Hence, the primary structure of this work is organized as follows:\\\\
In Section \ref{sec1}, we provide an overview of the current state of modern cosmology. This section emphasizes the various theoretical frameworks proposed to explain the Universe's late-time cosmic acceleration. Furthermore, we outline the objectives of our paper and the contributions it intends to make. Section \ref{sec2} offers a fundamental discussion of the essential equations in Horava-Lifshitz Gravity within the FLRW Universe. In Section \ref{sec3}, we employ two different parametrization approaches for the Equation of State (EoS) parameter in the model and calculate the corresponding Hubble solutions for each parametrization. Section \ref{sec4} contains a comprehensive explanation of the methodology employed in the manuscript to determine the best-fitting values for each model and to create 1D and 2D posterior distributions at confidence levels of $68.3\%$ (1$\sigma$) and $95.4\%$ (2$\sigma$) for each model. In Section \ref{sec5}, we compare our model predictions with the standard $\Lambda$CDM and CC datasets. Section \ref{sec6} presents graphical representations of cosmographic parameters, including deceleration and jerk. Section \ref{sec7} offers a detailed analysis of the $Om(z)$ diagnostic parameters. In Section \ref{sec8}, we present informative criteria for evaluating the model. Finally, in Section \ref{sec9}, we discuss the results, and we conclude our paper in Section \ref{sec10} with final remarks.

\section{Basic Equations in Horava-Lifshitz Gravity}\label{sec2}
In this section, we provide a concise overview of the scenario in which cosmological evolution is governed by HL gravity. The key dynamical fields associated with HL gravity are the lapse function ($N$), the shift functions ($N^i$), and the spatial metric ($g_{i j}$). The complete metric in terms of these basic variables is described by~\cite{calcagni2009cosmology,calcagni2010detailed,kiritsis2010spherically}

    \begin{equation}\label{HL1}
    ds^2=-N^2dt^2+g_{i j} \left(dx^i+N^i dt\right)\left(dx^j+N^j dt\right).
    \end{equation}
    where $i, j=1(1)3$ are raised and lowered using $g_{ij}$. The coordinate scaling transformation is expressed as $t\rightarrow l^3 t$ and $x^i\rightarrow l x^i$. Utilizing the projectability condition in conjunction with the detailed balanced principle, the gravity action can be expressed as \cite{hovrava2009membranes,hovrava2009quantum,lifshitz1941theory}

    \begin{equation}
    S=\int dt d^3 x \sqrt{g} N\left[\mathcal{L}_0 + \mathcal{L}_1 +\mathcal{L}_2\right]
    \end{equation}
    with
    \begin{eqnarray}
    \mathcal{L}_0&=&\frac{2\left(K_{ij}K^{ij}-\lambda K^2\right)}{\kappa^2}\\
    \mathcal{L}_1&=&\frac{\kappa^2 C_{ij}C^{ij}}{2\omega^4} -\frac{\kappa^2 \mu \epsilon^{i j k } R_{i, j} \Delta_j R^l_k}{2\omega^2 \sqrt{g}}+\frac{\kappa^2 \mu^2 R_{ij} R^{ij}}{8}\\
    \mathcal{L}_2&=&\frac{\kappa^2 \mu^2}{8(1-3\lambda)}\left\{\frac{(1-4\lambda)R^2}{4} +\Lambda R -3 \Lambda^2 \right\}
    \end{eqnarray}
    Here, $\lambda$, $\mu$, $\omega$ and $\kappa$ are dimensionless constant, $\Lambda$ is the cosmological constant and $R_{ij}$, $R$ are Ricci tensor and Ricci scalar respectively.
    Also, the extrinsic curvature and  the Cotton tensor are respectively defined as
    \begin{equation}
    K_{ij}=\frac{\dot{g}_{ij}-\Delta_i N_j-\Delta_j N_i}{2N}
    \end{equation}
    and
    \begin{equation}
     C^{ij}=\frac{\epsilon^{ijk} \Delta_k\left(R_i^j-\frac{R}{4} \delta^j_i\right)}{\sqrt{g}}
    \end{equation}
    In the above equations, all the covariant derivatives are determined with respect to the spatial metric $g_{ij} \epsilon^{ijk}$, a antisymmetric unit tensor.\\\\
    Now, considering the time-dependent lapse function $N\equiv N(t)$ and incorporating FRW metric with $N=1~,~g_{ij}=a^2(t)\gamma_{ij}~,~N^i=0$ with
    \begin{equation}
    \gamma_{ij}dx^i dx^j=\frac{dr^2}{1-kr^2}+r^2 d\Omega_2^2,
    \end{equation}
    the Friedmann equations by taking variation of $N$ and $g_{ij}$ are given by \cite{jamil2010new,paul2012modified}

    \begin{widetext}
    \begin{equation}\label{HLFriedmann1}
    H^2=\frac{\kappa^2\rho}{6\left(3\lambda-1\right)}+\frac{\kappa^2}{6\left(3\lambda-1\right)}\left[\frac{3\kappa^2\mu^2 k^2} {8\left(3\lambda-1\right)a^4}+\frac{3\kappa^2\mu^2 \Lambda^2}
    {8\left(3\lambda-1\right)}\right]-\frac{\kappa^4 \mu^2 \Lambda k}{8\left(3\lambda-1\right)^2a^2},
    \end{equation}

    \begin{equation}\label{HLFriedmann2}
    \dot{H}+\frac{3H^2}{2}=-\frac{\kappa^2 p}{4\left(3\lambda-1\right)} -\frac{\kappa^2} {4\left(3\lambda-1\right)}\left[\frac{3\kappa^2\mu^2 k^2} {8\left(3\lambda 1\right)a^4}+\frac{3\kappa^2\mu^2 \Lambda^2} {8\left(3\lambda-1\right)}\right]-\frac{\kappa^4 \mu^2 \Lambda k}{8\left(3\lambda-1\right)^2a^2}
    \end{equation}
    \end{widetext}
    where $a$ is the expansion factor, $H=\frac{\dot{a}}{a}$ is the usual Hubble parameter and $k$ is the curvature scalar corresponds to flat ($k=0$), open ($k=-1$) and closed ($k=1$) Universe. Also, $\rho$ and $p$ are the total energy density and total pressure of dark energy (DE) and dark matter (DM), respectively.\\\\
    Let us consider that the Universe is filled with DE and DM. Then, we have $\rho = \rho_m + \rho_d$ and $p =p_m + p_d$, respectively. Assuming separate conservation equations for DM and DE, we have

    \begin{equation}\label{DM}
    \dot{\rho}_m + 3H(\rho_m + p_m) = 0,
    \end{equation}
    and
    \begin{equation}\label{DE}
        \dot{\rho}_d + 3H(\rho_d + p_d) = 0.
    \end{equation}
    As DM is pressureless, i.e., $p_m = 0$, Equation \eqref{DM} yields $\rho_m = \rho_{m0}a^{-3}$. Let the equation of state parameter $w(z)=p/\rho$, so from equation \eqref{DE}, we
    obtain $\rho_d=\rho_{d0}~e^{3\int \frac{1+w(z)}{1+z} dz}$.
    Here, $\rho_{m0}$ and $\rho_{d0}$ are the present values of DM and DE energy densities, respectively. \\\\
    We can set, $G_{c}=\frac{\kappa^2}{16\pi \left(3\lambda-1\right)}$ with the condition $\frac{\kappa^4 \mu^2 \Lambda}{8\left(3\lambda-1\right)}=1$, so from detailed balance, the above Friedmann equations can be rewritten as
    \begin{equation}\label{H1}
    H^2=\frac{8\pi G_c}{3}\left(\rho_m +
    \rho_{d}\right)+\left(\frac{k^2} {2\Lambda
    a^4}+\frac{\Lambda}{2}\right)-\frac{k}{a^2},
    \end{equation}
    \begin{equation}\label{H2}
    \dot{H}+\frac{3}{2}H^2=-4\pi G_c p_d -\left(\frac{k^2}{4\Lambda
    a^4}+\frac{3\Lambda}{4}\right)-\frac{k}{2a^2}.
    \end{equation}
    Using the dimensionless parameters $\Omega_{i0}\equiv\frac{8\pi
    G_c}{3H_0^2}\rho_{i0}$, $\Omega_{k0}=-\frac{k}{H_0^2}$,
    $\Omega_{\Lambda 0}=\frac{\Lambda}{2H_0^2}$, we obtain
    \begin{widetext}
    \begin{equation}\label{E}
    H^2(z)=H_0^2\left[\Omega_{m0}(1+z)^3+\Omega_{k0}(1+z)^2
    +\Omega_{\Lambda 0}+\frac{\Omega_{k0}^2(1+z)^4}{4\Omega_{\Lambda
    0}}+\Omega_{d0}~e^{3\int \frac{1+w(z)}{1+z} dz}\right]
    \end{equation}
    with
    \begin{equation}
    \Omega_{m0}+\Omega_{d0}+\Omega_{k0}+\Omega_{\Lambda
    0}+\frac{\Omega_{k0}^2}{4\Omega_{\Lambda 0}}=1
    \end{equation}
    \end{widetext}

    \section{New Parametrizations of Dark Energy}\label{sec3}

    $\bullet$ {\bf Model I (CBDRM):} $``CBDRM"$
    (Chaudhary-Bouali-Debnath-Roy-Mustafa)-type parametrization is
    given by the EoS \cite{Chaudhary:2023vxz}
    \begin{equation}\label{New}
        w(z)=w_0+w_1 \frac{1+z}{2+z}
    \end{equation}
    where $\omega_0$, $\omega_1$ are constant parameter. It converges to a constant value, \(w_0 + w_1\), at high redshifts (\(z\rightarrow \infty\)), suggesting a consistent behavior in the distant past of the Universe. This implies that dark energy had a relatively stable and non-evolving equation of state in the early Universe described by \(w_0 + w_1\). Conversely, at low redshifts (\(z\rightarrow -1\)), the parametrization converges to \(w_0\), implying that in the present epoch, the dark energy equation of state tends to be described by a constant value \(w_0\).
    The energy density becomes
    \begin{equation}
        \rho_d=\rho_{d0}~(1+z)^{3(1+w_0)}(2+z)^{3w_{1}}
    \end{equation}
    So from equation (\ref{E}), we obtain
    
    \begin{widetext}
    \begin{eqnarray}
    H^2(z)= && H_0^2\left[\Omega_{m0}(1+z)^3+\Omega_{k0}(1+z)^2
    +\Omega_{\Lambda 0}+\frac{\Omega_{k0}^2(1+z)^4}{4\Omega_{\Lambda
    0}} \right.  \nonumber
    \\
    &&+\left. \left(1-\Omega_{m0}-\Omega_{k0}-\Omega_{\Lambda
    0}-\frac{\Omega_{k0}^2}{4\Omega_{\Lambda
    0}}\right)~(1+z)^{3(1+w_0)}(2+z)^{3w_{1}}\right]
    \end{eqnarray}
    \end{widetext}

    $\bullet$ {\bf Model II (CADMM):} $``CADMM"$
    (Chaudhary-Arora-Debnath-Mustafa-Maurya)-type parametrization is
    given by the EoS \cite{Chaudhary:2023ddn}

    \begin{equation}\label{New1}
    w(z)=w_0+\frac{\alpha+(1+z)^{\beta}}{w_{1}+w_{2}(1+z)^{\beta}}
    \end{equation}
    with constants \(w_0, w_1, w_2, \alpha, \beta\) exhibit both convergence and divergence behaviors. At high redshifts (\(z\rightarrow \infty\)), it converges to \(w_0 + \frac{1}{w_2}\), indicating a consistent behavior in the distant past of the Universe, given that \(w_0\), \(w_1\), \(w_2\), \(\alpha\), and \(\beta\) satisfy certain conditions. Conversely, as \(z\) approaches -1 (representing the present epoch), the behavior depends on parameter values and can either converge to \(w_0 + \frac{\alpha}{w_1}\) when \(\beta\) is greater than 0 or diverge when the limit is undefined or approaches a different value. These convergence and divergence characteristics underscore the sensitivity of this parametrization to the chosen parameters and highlight the importance of parameter selection for accurately describing dark energy's equation of state across cosmic epochs. The energy
    density becomes
    \begin{equation}
        \rho_d=\rho_{d0}~(1+z)^{3(1+w_0+\frac{\alpha}{w_{1}})}[w_{1}+w_{2}(1+z)^{\beta}]^{\frac{3(w_{1}-\alpha w_{2})}{\beta w_{1}w_{2}}}
    \end{equation}
    So from equation (\ref{E}), we obtain
    
    \begin{widetext}
        \begin{eqnarray}
    H^2(z)= && H_0^2\left[\Omega_{m0}(1+z)^3+\Omega_{k0}(1+z)^2
    +\Omega_{\Lambda 0}+\frac{\Omega_{k0}^2(1+z)^4}{4\Omega_{\Lambda
    0}} \right.  \nonumber
    \\
    &&+\left. \left(1-\Omega_{m0}-\Omega_{k0}-\Omega_{\Lambda
    0}-\frac{\Omega_{k0}^2}{4\Omega_{\Lambda
    0}}\right)~(1+z)^{3(1+w_0+\frac{\alpha}{w_{1}})}[w_{1}+w_{2}(1+z)^{\beta}]^{\frac{3(w_{1}-\alpha
    w_{2})}{\beta w_{1}w_{2}}}\right]
    \end{eqnarray}
    \end{widetext}
    In the case of the CBDRM Model, there are 5 depend on free parameters $\Omega_{mo}, \Omega_{Ko}, \Omega_{\Lambda0}, w_0$ and $w_1$. While there are 8 free parameters in CADMM Model, $\Omega_{mo}, \Omega_{Ko}, \Omega_{\Lambda0}, w_0$, $w_1$. $w_2$, $\alpha$ and $\beta$.\\\\

    \section{Methodology}\label{sec4}
    In our investigation of the late-time cosmic expansion of the Universe, we have employed a comprehensive dataset comprising the latest Baryon Acoustic Oscillation (BAO) measurements obtained from various observational experiments. These data points have been sourced from multiple sources, including the Sloan Digital Sky Survey (SDSS) \cite{42,43,44,45,46,49}, the Baryon Oscillation Spectroscopic Survey (BOSS) \cite{43}, and the extended Baryon Oscillation Spectroscopic Survey (eBOSS) \cite{45,46,47,48}. Additionally, our dataset incorporates measurements from other significant surveys such as the WiggleZ Dark Energy Survey \cite{41}, the Dark Energy Survey (DES) \cite{50}, the Dark Energy Camera Legacy Survey (DECaLS) \cite{44}, and the 6dF Galaxy Survey BAO (6dFGS BAO)\cite{44}. It's worth noting that one significant challenge associated with the BAO dataset is the potential presence of correlations among measurements obtained from different data releases. To address this issue and assess systematic errors, it is customary to employ mock datasets generated using N-body simulations with known cosmological parameters. These mock datasets enable us to derive the appropriate covariance matrices, aiding in the accurate analysis of the BAO measurements. Given that our analysis involves a combination of data from various experiments, it is important to acknowledge that we do not possess the exact covariance matrix detailing their interrelations, as this information is not readily available. In lieu of this, we have conducted a covariance analysis based on the methodology proposed in \cite{51}. For uncorrelated data points, the covariance matrix can be expressed as follows: $C_{i i}=\sigma_{i}^{2}$. To account for the potential influence of correlations among data points, we have introduced a specific number of non-diagonal elements into the covariance matrix while ensuring its symmetry. This method enables the introduction of positive correlations among up to twelve pairs of randomly selected data points, encompassing more than 68.8\% of the dataset. The positions of these non-diagonal elements are determined randomly, and the magnitude of these selected covariance matrix elements, denoted as $C_{i j}$, is set to $C_{i j}=0.5 \sigma_{i} \sigma_{j}$. Here, $\sigma_{i}$ and $\sigma_{j}$ represent the published $1 \sigma$ errors associated with data points $i$ and $j$, respectively. To constrain our cosmological model parameters, we augment the Baryon Acoustic Oscillations (BAO) dataset with thirty-one data points from the Cosmic Chronometers (CC) technique. This approach allows us to directly obtain the Hubble parameter at various redshifts, extending up to approximately $z \lesssim 2$. We select CC data because of its reliability; it primarily involves measuring the age difference between two galaxies that evolved passively and originated simultaneously but are slightly separated in redshift. CC data allows us to compute $\Delta z / \Delta t$, which is preferable to methods relying on absolute age determinations for galaxies \cite{72CC}. Our chosen CC data points are sourced from independent sources \cite{73CC,74CC,75CC,76CC,77CC,78CC,79CC}. Importantly, these references are not influenced by the Cepheid distance scale or any specific cosmological model. However, it's worth noting that they rely on stellar age modeling, established using robust stellar population synthesis techniques. For additional details, refer to \cite{75CC,77CC,80CC,82CC,81CC,83CC} for analyses related to CC systematics. We have also incorporated the Pantheon+ dataset, an updated version of the Pantheon dataset \cite{pantheon+} [pantheon+], which now includes 1701 data points related to Type Ia supernovae (SNIa). These SNIa observations span a redshift range from $0.001$ to $2.3$. It's worth noting that SNIa observations have been instrumental in uncovering the phenomenon of the Universe's accelerating expansion. Additionally, our investigation has been expanded to encompass 162 Gamma Ray Bursts (GRBs) \cite{demianski2017cosmology} within a redshift range of $1.44$ to $8.1$, as well as 24 observations of compact radio quasars \cite{roberts2017tests} spanning redshifts between $0.46$ and $2.76$. Furthermore, we have integrated the latest measurement of the Hubble constant, $H_{0}=73.04 \pm 1.04 (\mathrm{~km} / \mathrm{s}) / \mathrm{Mpc}$, as an additional prior referred to as R22 \cite{7}. In our analysis, we have employed a nested sampler, implemented within the open-source package Polychord \cite{55}, in conjunction with the GetDist package \cite{lewis2019getdist} for presenting the results.
    \begin{table*}
    \begin{center}
    \begin{tabular}{|c|c|c|c|c|c|c|}
    \hline
    \multicolumn{7}{|c|}{MCMC Results} \\
    \hline\hline
    Model & Parameters & Priors & BAO & BA0 + R22 & CC + SC + BAO & CC + SC + BAO + R22 \\[1ex]
    \hline
    & $H_0$ & [50,100] &$69.534787^{+5.074759}_{-6.763334}$ & $72.959488^{+1.081221}_{-2.262288}$ & $69.825421^{+1.191490}_{-2.172397}$ & $71.674089^{+0.734089 }_{-1.438451}$ \\[1ex]
    $\Lambda$CDM Model &$\Omega_{m0}$ &[0.,1.]  &$0.315787^{+0.024349}_{-0.063738}$ & $ 0.313245^{+0.026888}_{-0.062467}$ & $0.268223^{+0.013069}_{-0.031296}$ & $0.264769 ^{+0.013130}_{-0.029599}$   \\[1ex]
    &$\Omega_{\Lambda0}$  & [0.,1.] &$0.672428^{+0.020603}_{-0.035488}$  & $0.674885^{+0.022260}_{-0.042377}$ & $0.724841^{+0.009527}_{-0.019060}$ & $0.728206^{+0.009749}_{-0.016438}$\\[1ex]
    &$r_{d}$[Mpc]  & [100.,200.] &$148.615914_{-10.291069}^{+14.792347}$ & $141.207428_{-2.740041}^{+6.003713}$ &  $146.574936_{-2.588157}^{+5.229145}$ & $143.050380 _{-1.813570}^{+3.702038}$ \\[1ex]
    &$r_{d}/r_{fid}$  & [0.9,1.1] &$1.002369_{-0.071648}^{+0.097832}$ & $0.952378 _{-0.018130}^{+0.037983}$ &  $0.990380_{-0.018442}^{+0.036209}$ & $0.966154 _{-0.014653}^{+0.026010}$ \\[1ex]
    \hline
    & $H_0$ & [50,100] &$69.691456^{+4.884573}_{-7.342426}$ & $72.965882^{+1.017462}_{-1.877960}$ & $69.698993  ^{+1.958050}_{-3.782444}$ & $72.355058^{+1.004604}_{-1.920220}$ \\[1ex] 
    &$\Omega_{m0}$ &[0.25,0.35]  &$ 0.289729 ^{+0.019634}_{-0.032585}$ & $ 0.287776 ^{+0.019470}_{-0.032443}$ & $ 0.303047 ^{+0.020620}_{-0.037112}$ & $ 0.285878 ^{+ 0.014233}_{-0.026139}$   \\[1ex]
    CBDRM Model &$\Omega_{k0}$ &[-0.2,0.] &$-0.055482^{+0.044982}_{-0.091596}$  & $-0.051144^{+0.040369}_{-0.088691}$ & $-0.045215^{+0.033451}_{-0.068627}$ & $-0.057268^{+0.037010 }_{-0.078792}$ \\[1ex]
    &$\Omega_{\Lambda0}$  & [0.6,0.9] &$0.725861_{- 0.093745}^{+0.121922}$  & $0.725360  _{- 0.095621}^{+0.121447} $ & $0.753308 _{-0.093808 }^{+0.143736}$ & $ 0.667704 _{-0.049981}^{+0.065388}$ \\[1ex]
    &$w_{0}$  & [-1.,-0.4] &$-0.684491_{-0.216331}^{+0.301093}$ &$-0.675116_{-0.214116}^{+0.305689}$ & $-0.702575_{-0.193646}^{+0.281437}$ & $-0.694056_{-0.192856 }^{+0.278678}$ \\[1ex]
    &$w_{1}$  & [-0.5,0.] &$-0.238215_{-0.169197}^{+0.249623}$ & $-0.237544 _{-0.177217}^{+0.246685}$ &  $-0.276745_{-0.161080}^{+0.212444}$ & $-0.280404 _{-0.157553}^{+0.211559}$ \\[1ex]
    &$r_{d}$[Mpc]  & [100.,200.] &$148.881546_{-10.878719}^{+14.705531}$ & $141.972760 _{-3.852226}^{+7.711519}$ &  $146.584315_{-2.334845}^{+5.127246}$ & $144.835069 _{-2.378848}^{+4.426056}$ \\[1ex]
    &$r_{d}/r_{fid}$  & [0.9,1.1] &$1.004730_{-0.070653}^{+0.099196}$ & $0.957389 _{-0.027523}^{+0.049327}$ &  $0.989946_{-0.018378}^{+0.037511}$ & $0.978087 _{-0.016600}^{+0.032980}$ \\[1ex]
    
    \hline
    & $H_0$  & [50,80] &$68.758911^{+5.627074}_{-8.040622}$ & $72.847013^{+1.215592}_{-2.051350}$ & $70.115558 ^{+1.919837}_{-4.116930}$ & $72.347804^{+ 0.923328}_{-1.687843}$  \\[1ex]
    &$\Omega_{m0}$ &[0.,0.6]  &$ 0.268471 ^{+0.039748}_{-0.092760}$ & $0.250653 ^{+0.035346}_{-0.077143}$ & $0.276100 ^{+0.024887}_{-0.054324}$  & $0.269030 ^{+0.049190}_{-0.098134}$  \\[1ex]
    &$\Omega_{k0}$  &[-0.1,0.1] &$0.014332 ^{+0.065156}_{-0.101420}$  &  $0.020444^{+0.054223}_{-0.092289}$& $0.010925^{+0.059011}_{-0.095113}$ & $0.012041^{+0.062929}_{-0.074865}$ \\[1ex]
    &$\Omega_{\Lambda0}$  & [0.,1.] &$0.718196 _{ 0.256149}^{+0.428503}$  &$0.707604 _{- 0.215256}^{+0.426255} $ & $0.707827_{- 0.190084}^{+0.438503}$ & $0.725979 _{-0.174077}^{+0.306287}$ \\[1ex]
    CADMM Model &$w_{0}$ & [-3.,-2.] &$-2.529308_{-0.346948 }^{+0.449084}$ & $-2.514677_{-0.361487   }^{+0.457550}$ &$-2.598036_{-0.291603  }^{+0.384502}$ & $-2.464608_{-0.394155 }^{+0.515804}$  \\[1ex]
    &$w_{1}$  & [0.2,0.7] &$0.493478_{- 0.141291}^{+0.273149}$ & $0.488474 _{- 0.147085}^{+0.247643}$ & $0.484898_{-0.172273 }^{+0.260129}$& $0.504651_{-0.154310}^{+0.272959}$ \\[1ex]
    &$w_{2}$ & [0.2,1.] &$0.695235_{-0.199189 }^{+0.341742}$ & $0.667413_{-0.240303   }^{+0.316143}$ &$0.731066_{-0.213564  }^{+0.377010}$ & $0.672823_{-0.223094 }^{+0.306482}$  \\[1ex]
    &$\alpha$ & [0.5,1.5] &$1.048673_{-0.303737  }^{+0.504431}$ & $1.048785_{-0.296111   }^{+0.527336}$ &$0.935951_{-0.308283  }^{+0.416779}$ & $1.044758 _{-0.391627 }^{+0.506546}$  \\[1ex]
    &$\beta$ & [1.,3.] &$1.944875_{-0.658647 }^{+0.900190}$ & $2.014418_{-0.740012   }^{+0.982176}$ &$1.873710_{-0.650020  }^{+0.838315}$ & $2.150797_{-0.715336 }^{+1.057201}$  \\[1ex]
    &$r_{d}$[Mpc]  & [100,200] &$148.463181_{-10.580674 }^{+16.256941}$ & $140.816016 _{-4.013211}^{+6.606581}$ &  $146.904120_{-2.356544}^{+4.546238}$ & $144.466836 _{-2.421153}^{+4.288758}$ \\[1ex]
    &$r_{d}/r_{fid}$  & [0.9,1.1] &$1.007496_{-0.074382}^{+0.099445}$ & $0.949938 _{-0.027541}^{+0.045378}$ &  $0.990363_{-0.017981}^{+0.032661}$ & $0.974860 _{-0.014041}^{+0.030990}$ \\[1ex]
    \hline
    \end{tabular}
    \caption{The table provides constraints on cosmological parameters for the $\Lambda$CDM, CBDRM, and CADMM models. The data sources include cosmic chronometers (CC). Additionally, the combination of three datasets, namely Type Ia supernovae (SNIa), quasar measurements (Q), and Gamma Ray Bursts (GRB), is represented by SC. We incorporate the latest Hubble constant measurement, denoted as R22, at a 95\% confidence level (CL).}\label{tab_MCMC}
    \label{tab_2}
    \end{center}
    \end{table*}
    \begin{figure*}
    \centering
    \includegraphics[scale=0.6]{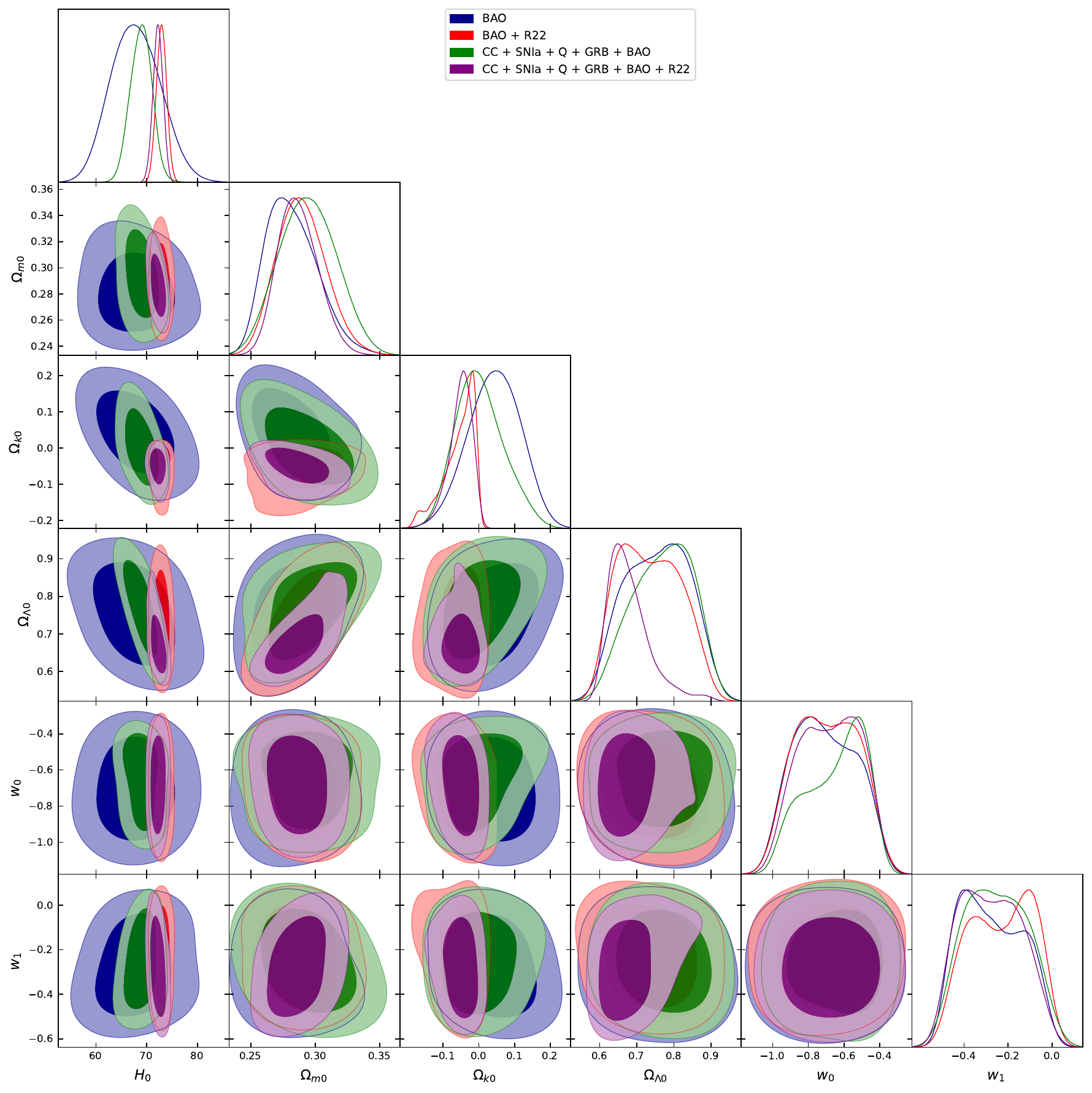}
    \caption{The figure illustrates the posterior distribution for various observational datasets of the CBDRM Model, with 1$\sigma$ and 2$\sigma$ confidence levels. In this context (CC) represents Cosmic Chronometers,(SNIa) refers to Type Ia supernovae, measurements (Q) corresponds to quasars, (GRB)  stands for Gamma Ray Bursts, and R22 denotes the latest Hubble constant measurement,}\label{fig_2}
    \end{figure*}
    \begin{figure*}
    \centering
    \includegraphics[scale=0.4]{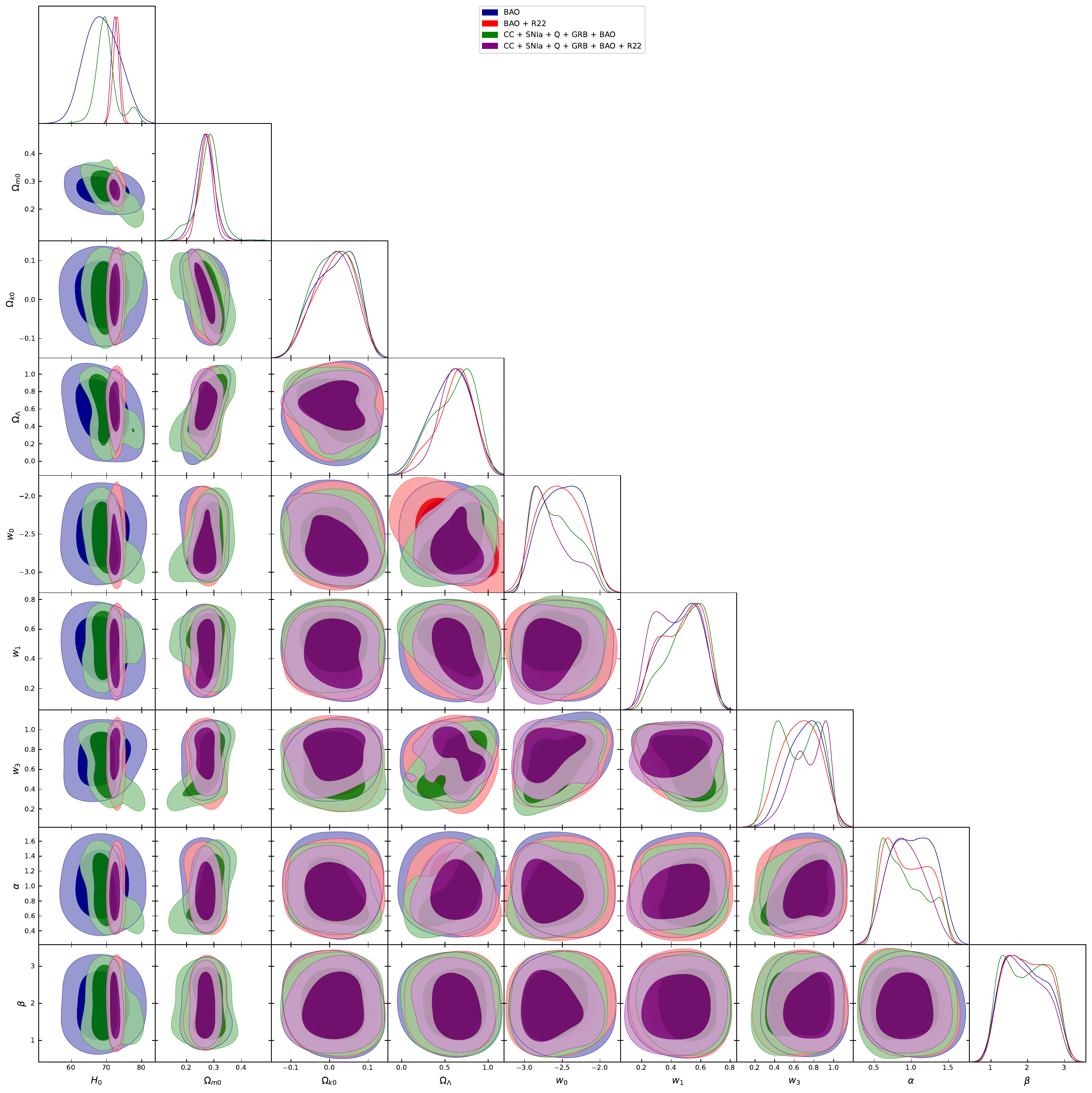}
    \caption{The figure illustrates the posterior distribution for various observational datasets of the CADMM Model, with 1$\sigma$ and 2$\sigma$ confidence levels. In this context (CC) represents Cosmic Chronometers,(SNIa) refers to Type Ia supernovae, measurements (Q) corresponds to quasars, (GRB)  stands for Gamma Ray Bursts, and R22 denotes the latest Hubble constant measurement,}\label{fig_3}
    \end{figure*}
    \clearpage
    \section{Observational and theoretical comparisons of the Hubble Function}\label{sec5}
    After finding the best parameter values for the CBDRM and CADMM models ( see Table \ref{tab_2} ), it's important to compare them with the well-established $\Lambda$CDM. The $\Lambda$CDM model has consistently proven to work well with various observational data, making it a strong framework for understanding the Universe's evolution. This comparison helps us understand the differences between our models and the widely accepted $\Lambda$CDM model, giving us insights into what makes our parametrized models unique, especially in terms of how the Universe behaves. By examining how our models deviate from $\Lambda$CDM, We can pinpoint distinct characteristics that set them apart. This examination yields valuable insights into both the advantages and limitations of our models, 
    \subsection{Comparison with the CC data points}
    We conducted a comparative analysis of the CBDRM and CADMM cosmological models, using observational data obtained from Cosmic Chronometers (CC). Additionally, we included the well-established $\Lambda$CDM model in our study for reference.
    \begin{figure}
    \includegraphics[scale=0.4]{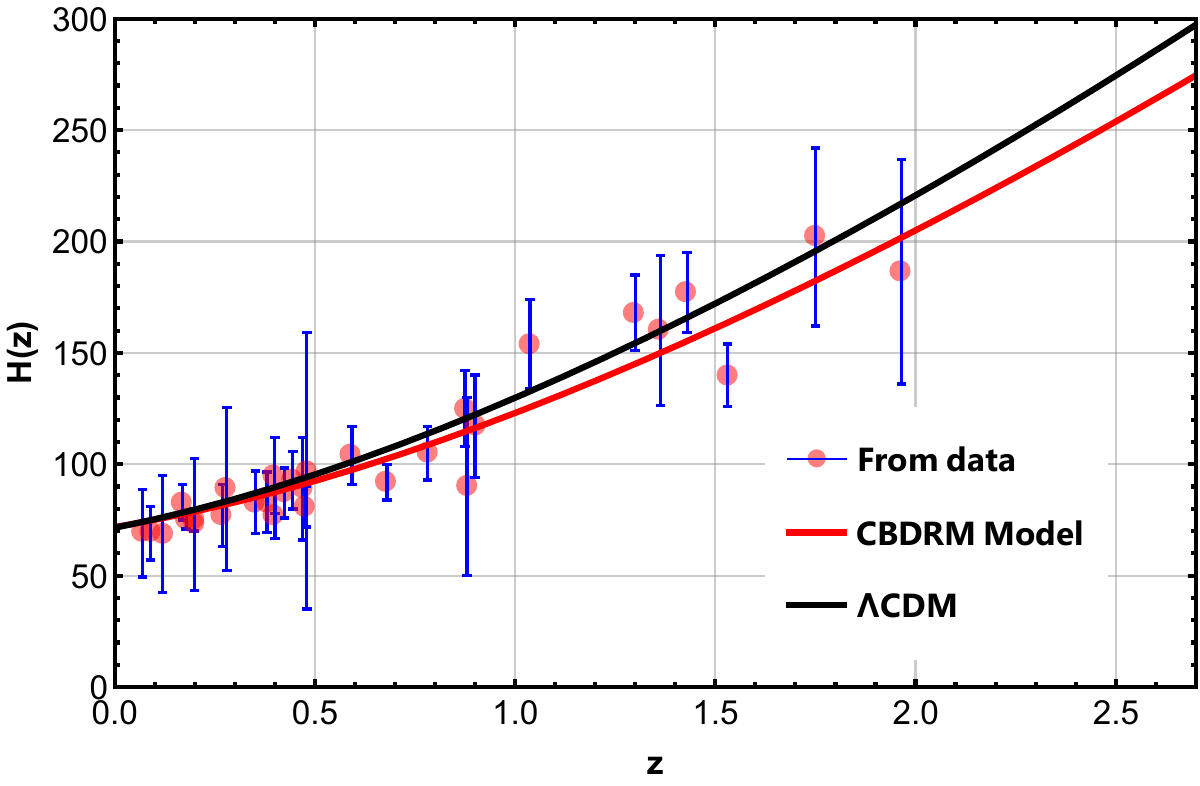}
    \caption{Theoretical curve of Hubble function $H(z)$ of the CBDRM model shown in red line and $ \Lambda$CDM model shown in black line with $\Omega_{\mathrm{m0}}=$ 0.3 and $\Omega_\Lambda =$ 0.7, against CC measurements shown in magenta dots with their corresponding error bars in blue line.}\label{fig_4}
    \end{figure}
    \begin{figure}
    \includegraphics[scale=0.4]{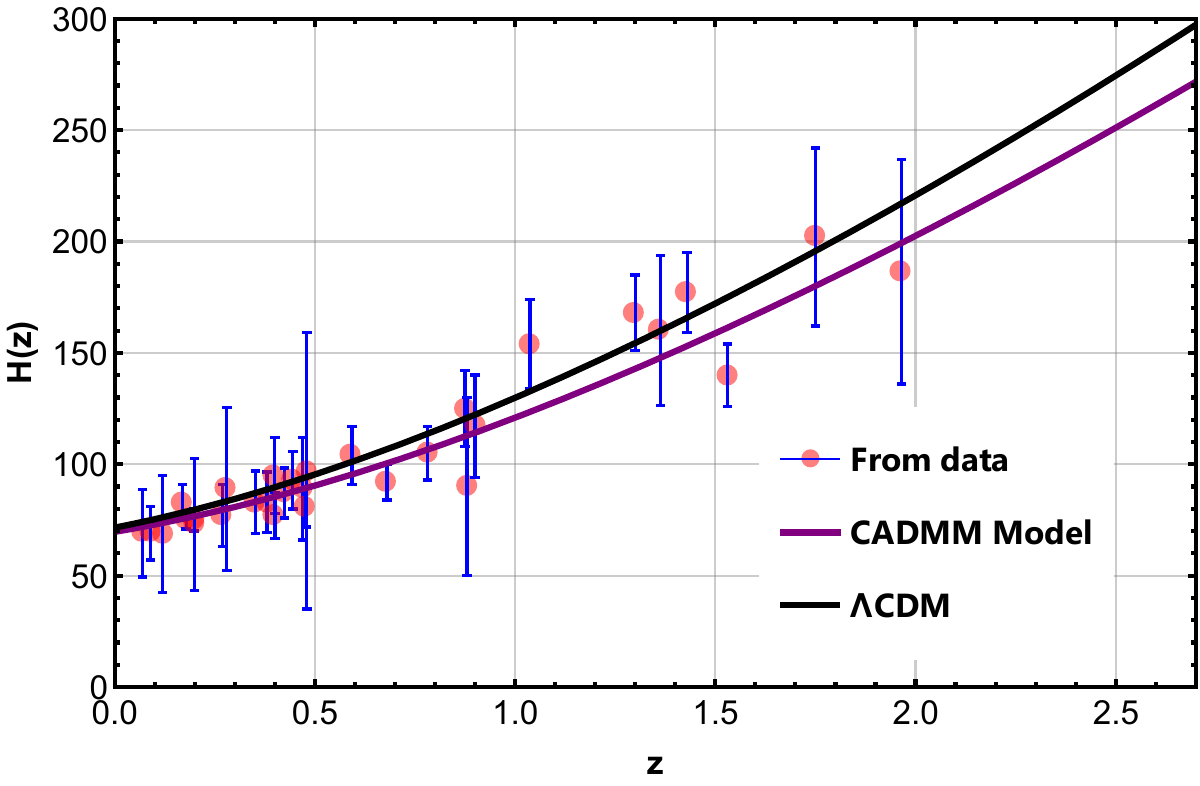}
    \caption{Theoretical curve of Hubble function $H(z)$ of the CADMM model shown in purple line and $ \Lambda$CDM model shown in black line with $\Omega_{\mathrm{m0}}=$ 0.3 and $\Omega_\Lambda =$ 0.7, against CC measurements shown in magenta dots with their corresponding error bars in blue line.}\label{fig_5}
    \end{figure}
    \subsection{Relative difference between model and $\Lambda$CDM}
    We carried out a detailed comparison between the CBDRM and CADMM models and the conventional $\Lambda$CDM model to examine their distinct evolutionary behaviors at low and high redshifts..
    \begin{figure}
    \includegraphics[scale=0.42]{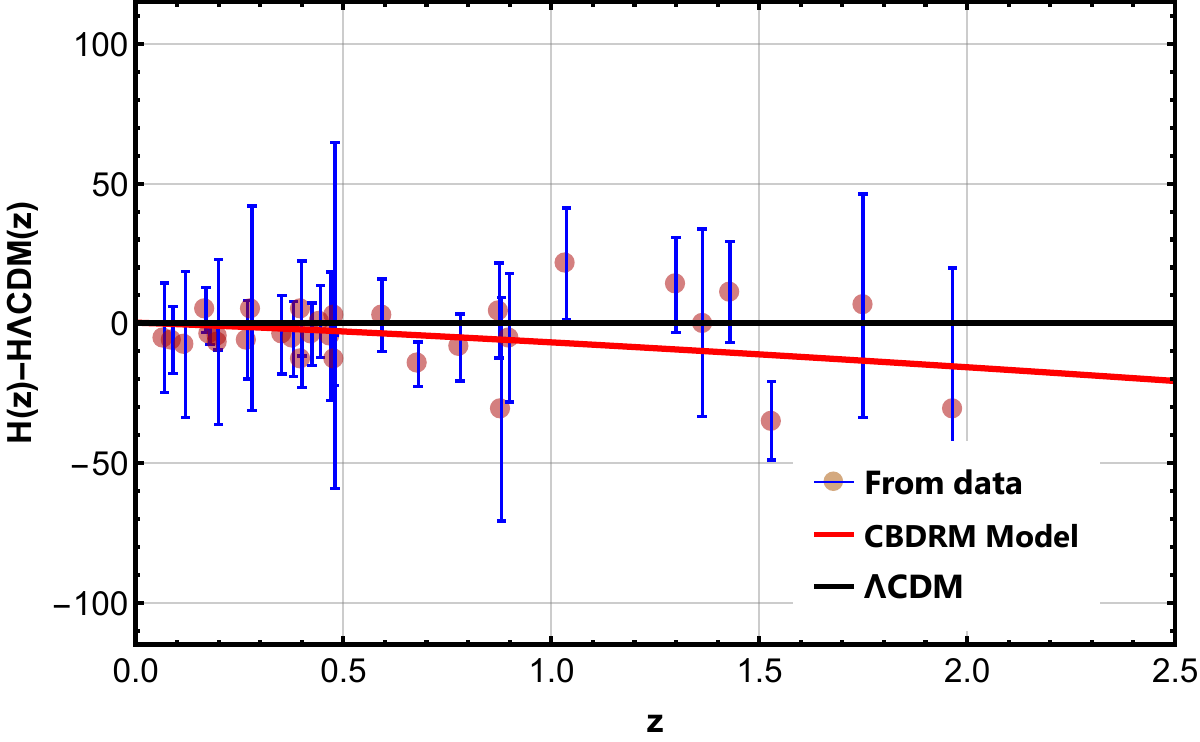}
    \caption{The variations of the difference between the CBDRM Model Hubble function shown in the red line and the $\Lambda$CDM Hubble function shown in the black line as a function of the redshift $z$ with $\Omega_{\mathrm{m0}}=$ 0.3 and $\Omega_\Lambda =$ 0.7, against CC measurements are shown in Magenta dots with their corresponding error bars shown in the blue line.}\label{fig_6}
    \end{figure}
    \begin{figure}
        \includegraphics[scale=0.42]{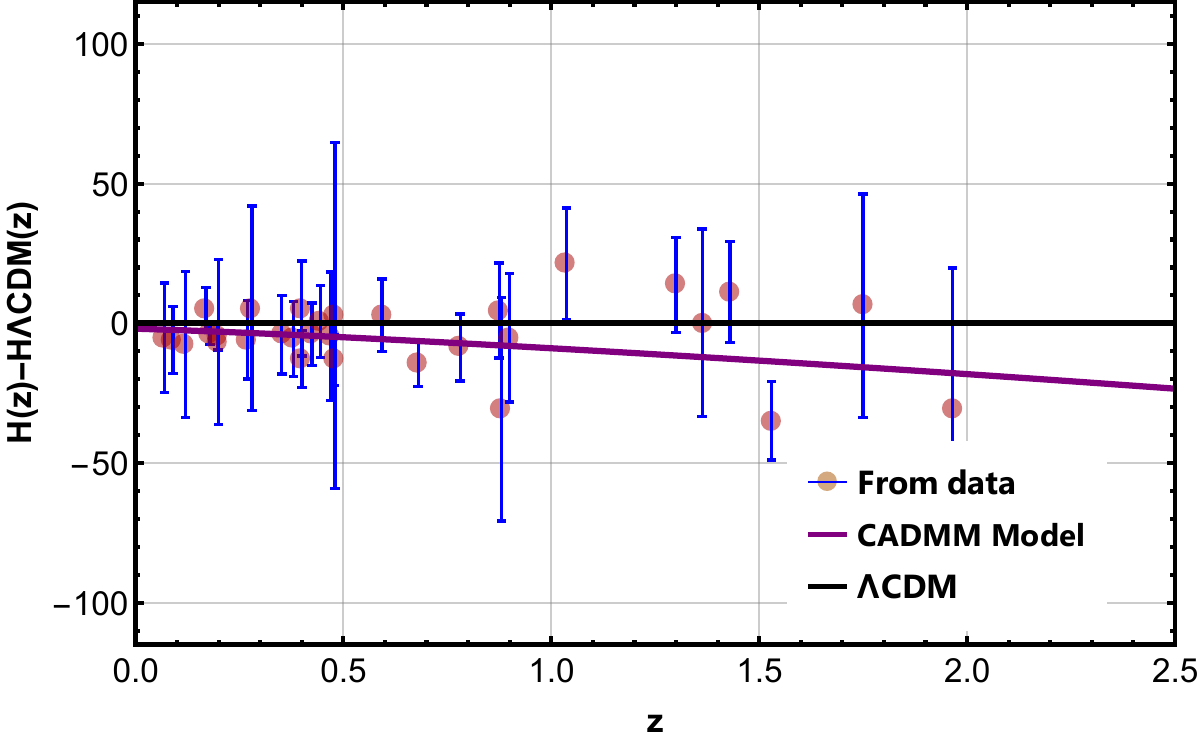}
    \caption{The variations of the difference between the CBDMM Model Hubble function shown in the purple line and the $\Lambda$CDM Hubble function shown in the black line as a function of the redshift $z$ with $\Omega_{\mathrm{m0}}=$ 0.3 and $\Omega_\Lambda =$ 0.7, against CC measurements are shown in Magenta dots with their corresponding error bars shown in the blue line.}\label{fig_7}
    \end{figure}
\section{Cosmography Parameters}\label{sec6}
Cosmography, originally introduced by Visser in his work \cite{visser2005cosmography}, occupies a central role in modern cosmology. It stands as a foundational tool for unraveling the mysteries surrounding the expansion of the Universe. In our research article, we leverage the powerful concept of cosmography to deepen our understanding of cosmic dynamics. We achieve this by meticulously analyzing observational data and comparing it to various theoretical models. These models encompass the CBDRM and CADMM models, alongside the well-established $\Lambda$CDM paradigm. By adopting this approach, we gain invaluable insights into how the Universe evolves across different redshifts. Cosmography equips us with the means to explore the cosmos across various temporal epochs, making it an indispensable instrument for advancing our comprehension of the Universe's past, present, and future.
    
\subsection{The deceleration parameter}
The deceleration parameter, often denoted as "q" and introduced by Edwin Hubble in the early 20th century, is a fundamental cosmological parameter crucial for the study of the Universe's expansion dynamics. Mathematically, it is defined as:
\begin{equation}
  q = -\frac{a\ddot{a}}{\dot{a}^2},   
\end{equation}
Where "a(t)" represents the Universe's scale factor as a function of time, "a dot" ("$\dot{a}$) signifies its first derivative, and "a double dot" ("$\ddot{a}$") is the second derivative. The deceleration parameter offers significant insights into the historical and future evolution of our cosmic environment. When the deceleration parameter is positive, it indicates a gradual slowdown in the Universe's expansion. A value of zero suggests that the expansion remains at a constant rate, often referred to as a "critical Universe." Conversely, a negative deceleration parameter points to the acceleration of cosmic expansion. This discovery became particularly prominent in the latter part of the 20th century when dark energy was found, providing a compelling explanation for the observed acceleration of cosmic expansion \cite{hubble1929relation}. In modern cosmology, the study of the deceleration parameter has gained increasing importance, especially in unraveling the mysteries of dark energy and the ultimate fate of the Universe. It serves as a crucial tool in observational cosmology, enabling us to investigate the properties of cosmic components like dark matter and dark energy, while also shedding light on the overall geometry of the Universe. This is an essential element in our quest to understand the cosmos.
\begin{figure}
\includegraphics[scale=0.42]{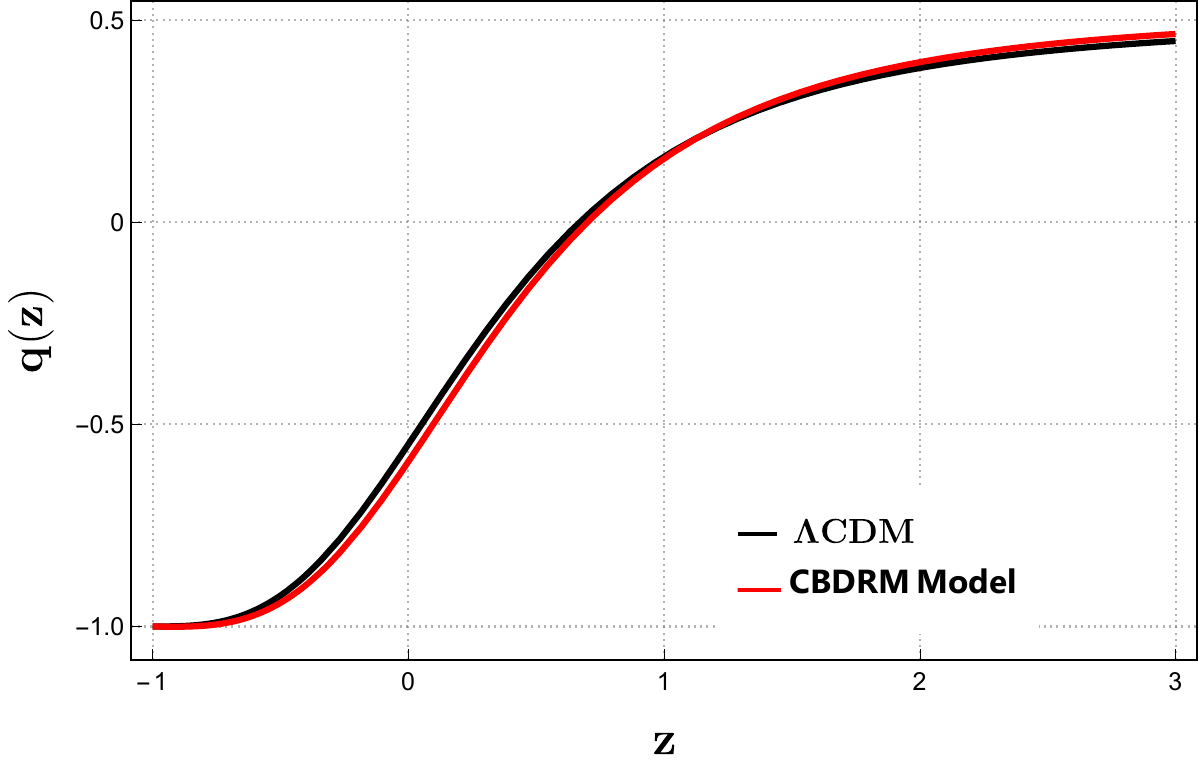}
\caption{The figure illustrates the evolution of the deceleration parameter concerning redshift ($z$). It includes the $\Lambda$CDM model, denoted by the Black ine, characterized by $\Omega_{\mathrm{m0}}=0.3$ and $\Omega_\Lambda=0.7$. Additionally, it showcases the CBDRM Model represented by the red line, utilizing the best-fit values derived from the CC + SC + BAO analysis.}\label{fig_8}
\end{figure}
\begin{figure}
\includegraphics[scale=0.42]{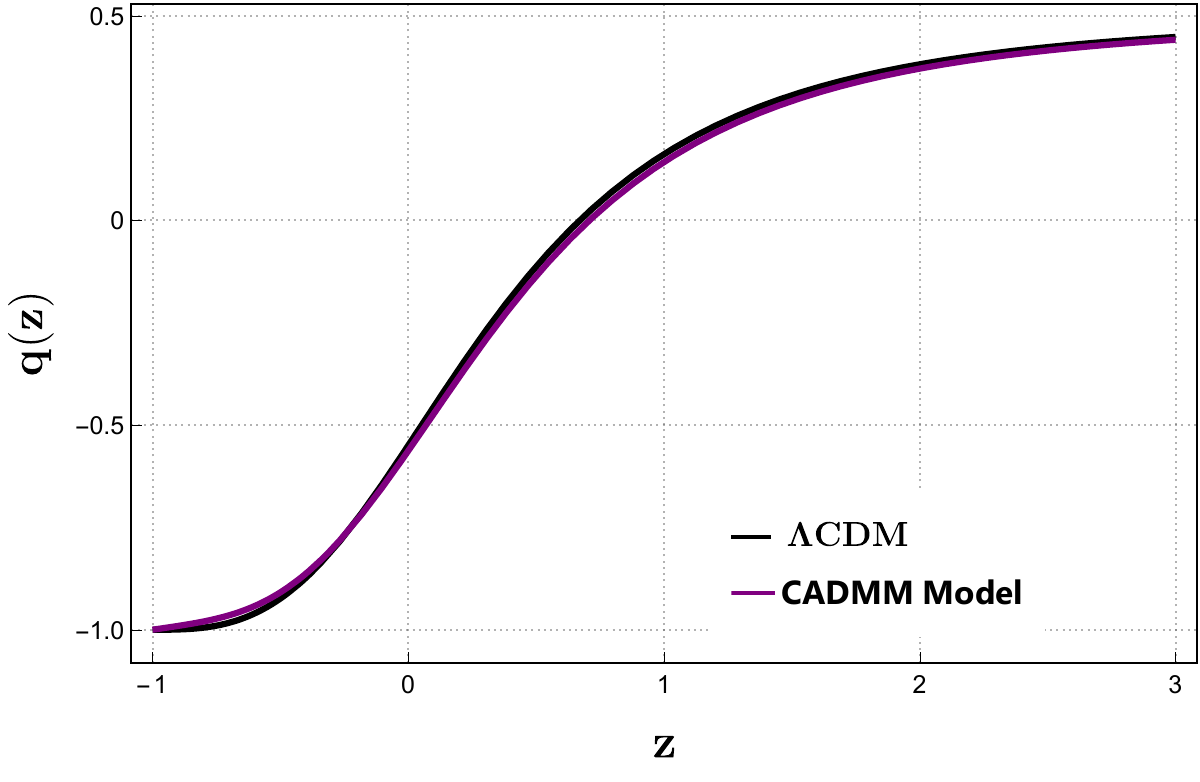}
\caption{The figure illustrates the evolution of the deceleration parameter concerning redshift ($z$). It includes the $\Lambda$CDM model, denoted by the Black ine, characterized by $\Omega_{\mathrm{m0}}=0.3$ and $\Omega_\Lambda=0.7$. Additionally, it showcases the CADMM Model represented by the purple line, utilizing the best-fit values derived from the CC + SC + BAO analysis.}\label{fig_9}
\end{figure}
\subsection{The jerk parameter}
In the field of cosmology, a fundamental objective is to gain a profound understanding of the Universe's dynamics. While the Hubble constant and the deceleration parameter have been pivotal in characterizing cosmic expansion, a more intricate parameter known as the "jerk parameter" has emerged, enriching our cosmological toolkit \cite{visser2004jerk}. The jerk parameter, denoted as "$j$," offers deeper insights into cosmic acceleration, complementing the information provided by the deceleration parameter. It represents the third time derivative of the Universe's scale factor, building upon the foundational concepts embodied by the Hubble parameter and the deceleration parameter. This parameter can be expressed mathematically as:
\begin{equation}
j = \frac{1}{a}\frac{d^3a}{d\tau^3}\left[\frac{1}{a}\frac{da}{d\tau}\right]^{-3} = q(2q + 1) + (1 + z)\frac{dq}{dz},    
\end{equation}
In this equation, \(j\) symbolizes the jerk parameter, \(a(t)\) represents the time-dependent scale factor of the Universe, and \(\dot{a}\), \(\ddot{a}\), and \(\dddot{a}\) signify the first, second, and third derivatives of the scale factor, respectively. The variable \(z\) denotes redshift.\\\\\
    \begin{figure}
    \includegraphics[scale=0.42]{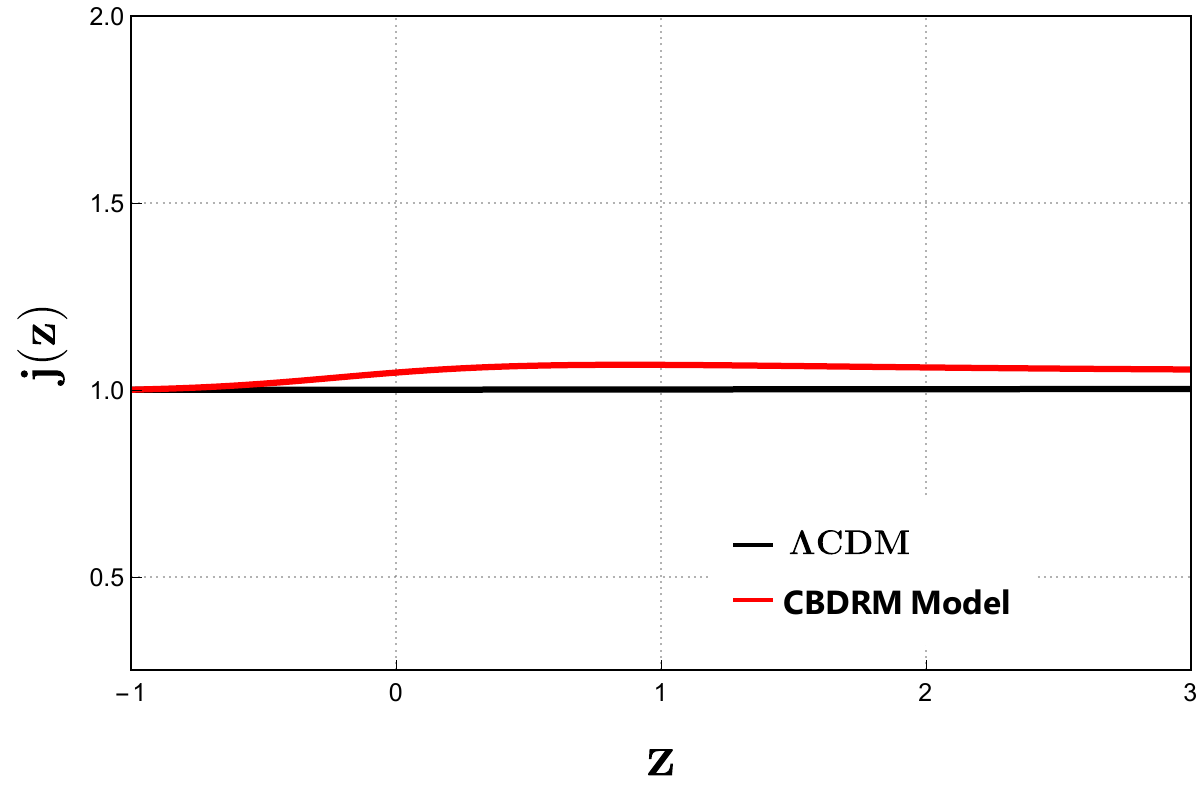}
    \caption{The figure illustrates the evolution of the jerk parameter concerning redshift ($z$). It includes the $\Lambda$CDM model, denoted by the Black ine, characterized by $\Omega_{\mathrm{m0}}=0.3$ and $\Omega_\Lambda=0.7$. Additionally, it showcases the CBDRM Model represented by the red line, utilizing the best-fit values derived from the CC + SC + BAO analysis.}\label{fig_10}
    \end{figure}
    \begin{figure}
    \includegraphics[scale=0.42]{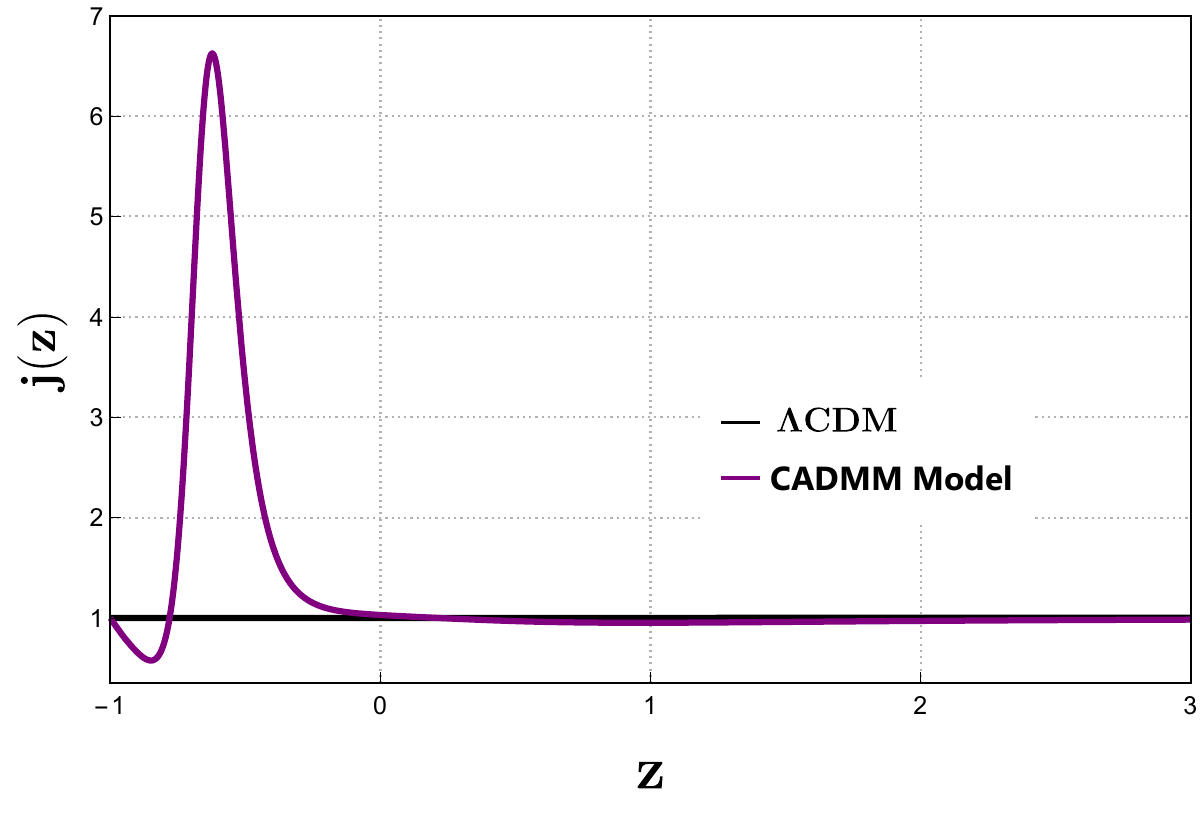}
    \caption{The figure illustrates the evolution of the jerk parameter concerning redshift ($z$). It includes the $\Lambda$CDM model, denoted by the Black ine, characterized by $\Omega_{\mathrm{m0}}=0.3$ and $\Omega_\Lambda=0.7$. Additionally, it showcases the CADMM Model represented by the purple line, utilizing the best-fit values derived from the CC + SC + BAO analysis.}\label{fig_11}
    \end{figure}
\section{Om Diagnostic}\label{sec7}
The $Om$ diagnostic is a powerful tool in cosmology used to distinguish different dark energy models from the standard $\Lambda$CDM model \cite{Om1,Om2,Om3,Om4}. It focuses on the equation of state ($w$) of dark energy and provides a straightforward way to categorize these models: For a flat $\Lambda$CDM model (where $w_0 = -1$), the $Om(z)$ parameter simplifies to $Om(z) = \Omega_{m0}$. This reflects a constant $Om(z)$ value. In the case of quintessence models (where $w_0 > -1$), the $Om(z)$ parameter exceeds the matter density at a given redshift ($\Omega_{m0}$). This results in a clear positive shift in the $Om(z)$ value. Conversely, for phantom models (where $w_0 < -1$), the $Om(z)$ parameter falls below the matter density ($\Omega_{m0}$), leading to a negative shift in the $Om(z)$ value \citep{escamilla2016nonparametric}. In a flat Universe, $Om(z)$ is defined as follows:
\begin{equation}
Om(z)=\frac{\left( \frac{H(z)}{H_{0}}\right) ^{2}-1}{(1+z)^{3}-1}\text{.}\label{34}
\end{equation}

    \begin{figure}
    \includegraphics[scale=0.42]{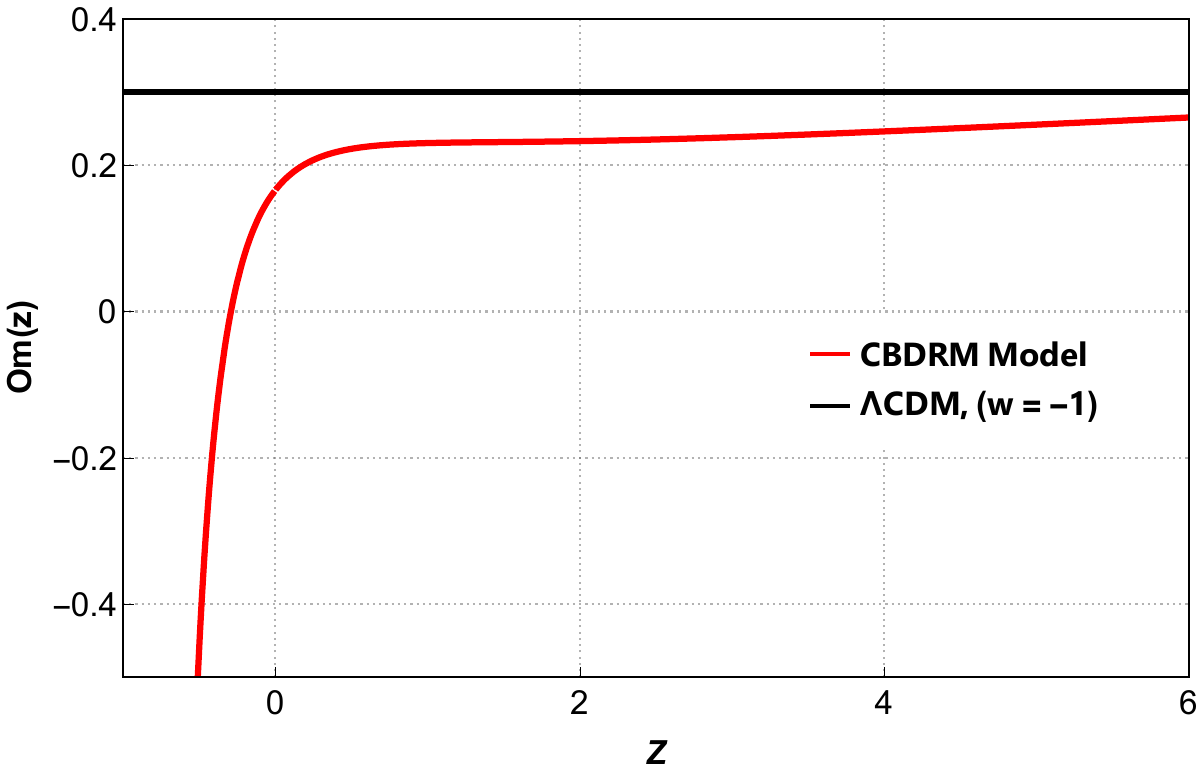}
    \caption{The figure depicts the $Om(z)$ profile of the CBDRM Model, represented by the red line, using the best-fit values derived from the CC + SC + BAO  analysis.}\label{fig_12}
    \end{figure}
    \begin{figure}
    \includegraphics[scale=0.42]{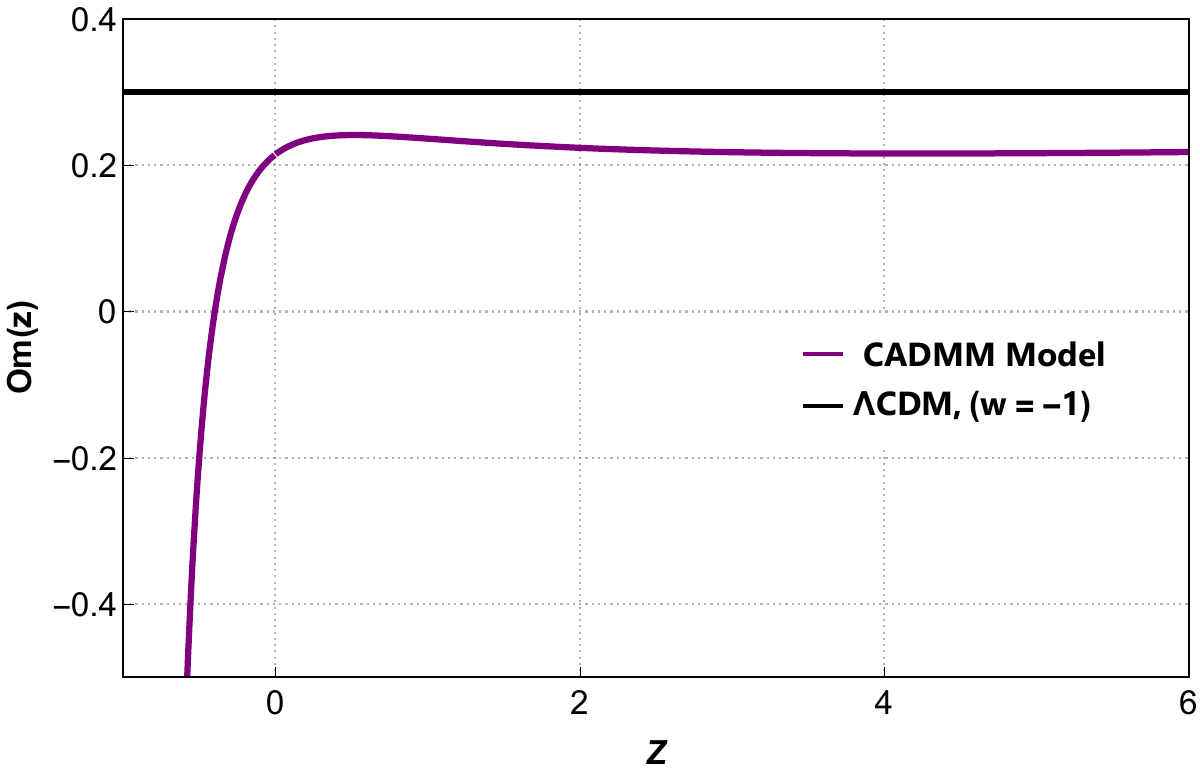}
    \caption{The figure depicts the $Om(z)$ profile of the CADMM Model, represented by the purple line, using the best-fit values derived from the CC + SC + BAO  analysis.}\label{fig_13}
    \end{figure}
    \section{$r_{d}$ vs $H_{0}$ Contour Plane}\label{sec8}
    The BAO scale is defined by a significant cosmic event referred to as the drag epoch, represented by the symbol $z_{d}$. This epoch signifies the point at which baryons and photons, initially tightly coupled, undergo a separation process, resulting in the emergence of a distinctive scale within the Universe. This scale's mathematical expression can be described as follows:
    \begin{equation}
    r_{d} = \int_{z_{d}}^{\infty} \frac{c_{s}(z)}{H(z)} \mathrm{d} z  \end{equation}
    Here, the speed of sound, denoted as $c_{s}$, is defined as:
    \begin{equation}
    c_{s} = \sqrt{\frac{\delta p_{\gamma}}{\delta \rho_{B}+\delta \rho_{\gamma}}} = \sqrt{\frac{(1 / 3) \delta \rho_{\gamma}}{\delta \rho_{B}+\delta \rho_{\gamma}}} = \frac{1}{\sqrt{3(1+R)}}
    \end{equation}
    In this context, $R$ represents the ratio of perturbations in baryonic matter density ($\delta \rho_{B}$) to those in photon density ($\delta \rho_{\gamma}$). Mathematically, it is given by $R \equiv \delta \rho_{B} / \delta \rho_{\gamma} = \frac{3 \rho_{B}}{4 \rho_{\gamma}}$.
    The data obtained from \cite{1} provides a precise measurement of the redshift at the drag epoch, denoted as $z_{d}$, with a value of $1059.94 \pm 0.30$. In the context of a flat $\Lambda$CDM model, the measurements from \cite{1} yield an estimation of the sound horizon, $r_{d}$, equal to $147.09 \pm 0.26$ megaparsecs (Mpc). Additionally, \cite{60} reports a sound horizon estimation of $143.9 \pm 3.1$ Mpc. Furthermore, \cite{63} employs Binning and Gaussian methods to combine 2D BAO and Type Ia Supernova (SNIe) data, resulting in absolute BAO scale values ranging from $141.45$ Mpc to $159.44$ Mpc (Binning) and from $143.35$ Mpc to $161.59$ Mpc (Gaussian). These diverse measurements and analyses underscore the precision and variability in estimating the sound horizon $r_{d}$, highlighting the ongoing efforts to refine our understanding of this fundamental cosmological parameter. The above findings reveal a significant disparity between observational measurements conducted during different cosmic epochs, akin to the well-known $H_{0}$ tension. It's important to note that our results are sensitive to the chosen range of priors for both $r_{d}$ and $H_{0}$, leading to shifts in the estimated values within the $r_{d}-H_{0}$ contour plane. The results depend critically on the priors for $r_{d}$ and $H_{0}$. From our numerical experiments, it is clear that using a tighter prior for $r_{d}$ will also move $H_{0}$ to a value compatible with that $r_{d}$, i.e. it will move it in the $r_{d}-H_{0}$ plane. This means that the effect from the choice of a prior for $r_{d}$ may be as strong as using a tight prior such as R22 on $H_{0}$. On the other hand, the tight prior on $H_{0}$ combined with a large interval for $r_{d}$ essentially swings the results towards the $H_{0}$ measurement on which the prior is centered. A noteworthy observation is that, in the absence of additional priors, our results tend to align remarkably well with the findings of reference \cite{1} and the results from the SDSS. This alignment is particularly evident when considering the BAO dataset alone, and it also corresponds closely to the results reported in \cite{63}. This sensitivity to prior choices underscores the intricate nature of cosmological parameter estimation and the importance of careful consideration when interpreting and comparing results. It highlights the ongoing effort to reconcile and understand the observed tensions between early and late-time cosmological measurements.\\\\
    \begin{figure}
    \centering
    \includegraphics[scale=0.42]{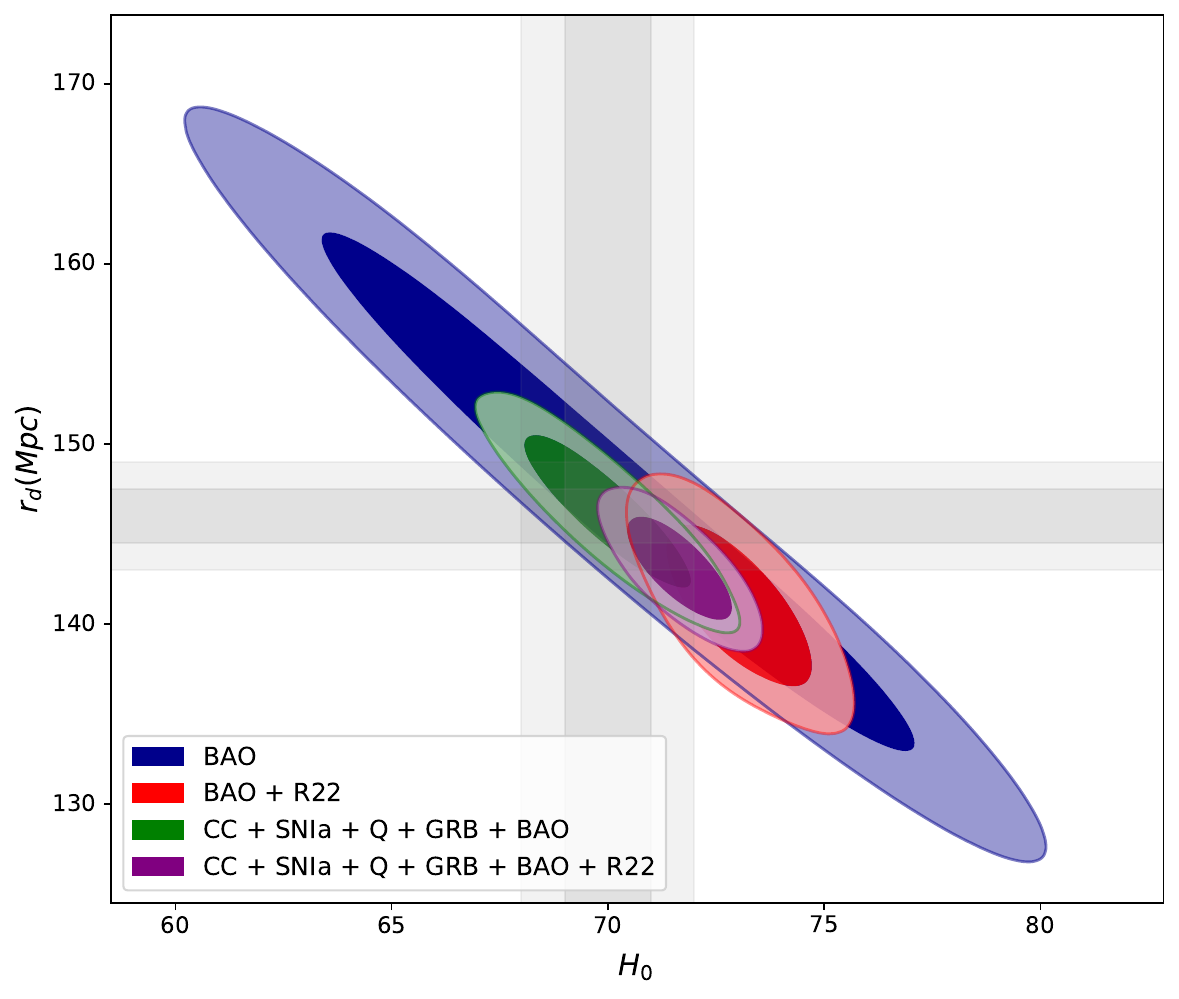}
    \caption{The figure illustrates the posterior distribution of diverse observational data measurements within the $r_{d}$ vs $H_{0}$ contour plane using the $\Lambda$CDM model. The shaded regions correspond to the 1$\sigma$ and 2$\sigma$ confidence plane.}\label{fig_14}
    \end{figure}
    \begin{figure}
    \centering
    \includegraphics[scale=0.38]{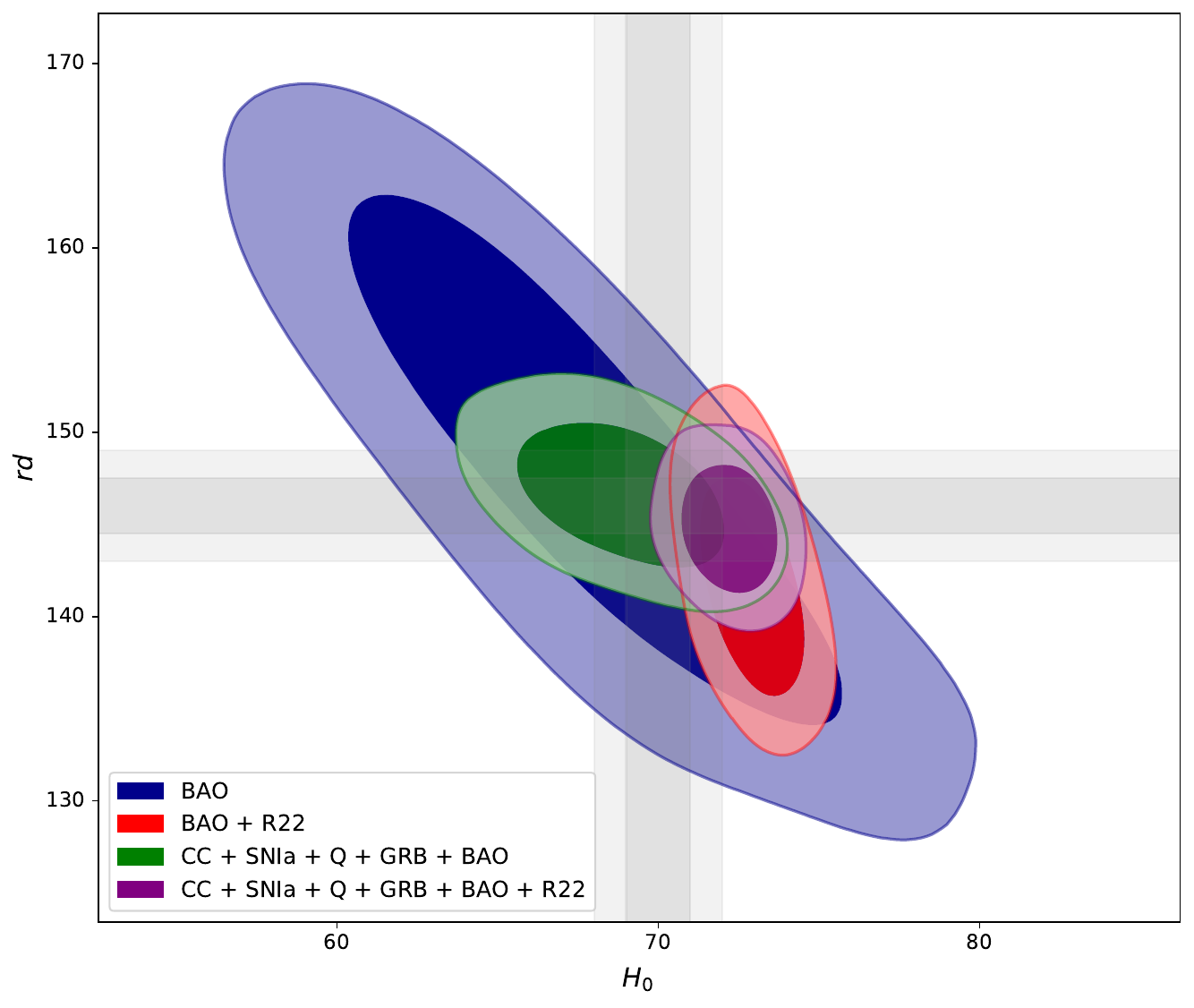}
    \caption{The figure illustrates the posterior distribution of diverse observational data measurements within the $r_{d}$ vs $H_{0}$ contour plane using the CBDRM model. The shaded regions correspond to the 1$\sigma$ and 2$\sigma$ confidence plane.}\label{fig_15}
    \end{figure}
    \begin{figure}
    \centering
    \includegraphics[scale=0.38]{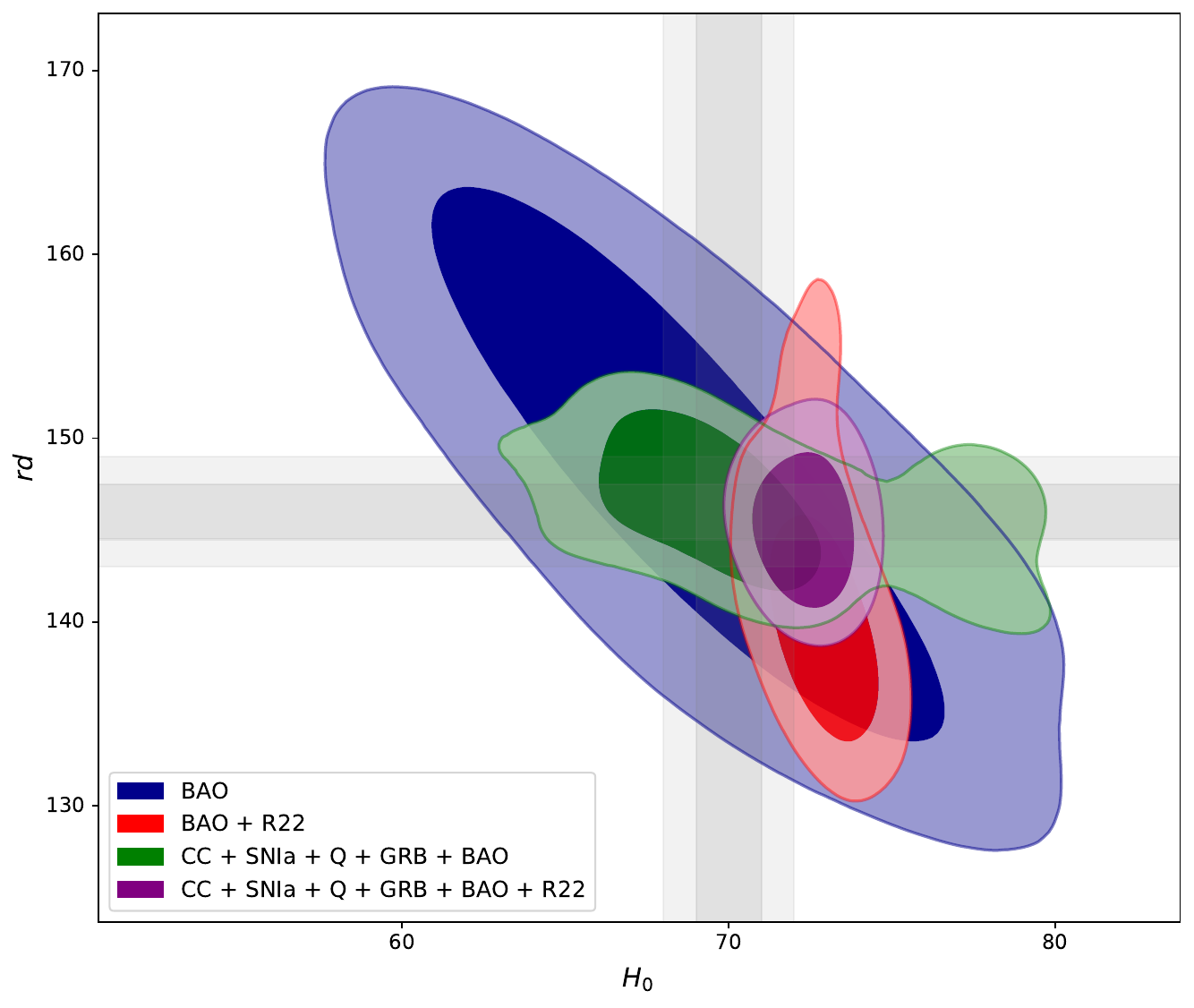}
    \caption{The figure illustrates the posterior distribution of diverse observational data measurements within the $r_{d}$ vs $H_{0}$ contour plane using the CADMM model. The shaded regions correspond to the 1$\sigma$ and 2$\sigma$ confidence plane.}\label{fig_16}
    \end{figure}.

\section{Information Criteria}\label{sec9}
To perform a statistical comparison between the CBDRM, CADMM, and $\Lambda$CDM models, we employ well-established information criteria, namely the Akaike Information Criterion (AIC) and the Bayesian Information Criterion (BIC) \cite{64,65,66}. These criteria help us assess the relative goodness-of-fit of different models, taking into account both their ability to describe the data and their complexity. The AIC is calculated using the formula: $\text{AIC} = \chi_{\text{min}}^{2} + 2k,$, where $\chi_{\text{min}}^{2}$ is the minimum chi-squared value achieved by the model during fitting, and $k$ represents the number of free parameters in the model. In our analysis, we have $N = 1305$ data points. The BIC is calculated as: $\text{BIC} = \chi_{\text{min}}^{2} + k \ln(N),$.  where $\ln(N)$ accounts for the number of data points. With these formulas, we can calculate the AIC and BIC for both the standard $\Lambda$CDM model and the alternative CADMM and CBDRM models.
\begin{table*}
\begin{center}
{\begin{tabular}{|c|c|c|c|c|c|c|c|c|}
\hline
Model & ${\chi_{\text{tot}}^2}^{min} $ & $\chi_{\text {red }}^2$ &$\mathcal{K}_{\textrm{f}}$ & $A I C_c$ & $\Delta A I C$ &{\bf$B I C$} & {\bf $\Delta B I C$}  \\[0.1cm]
\hline
$\Lambda$CDM Model & 1756.22  & 0.9867 & 3& 1762.22 & 0 & 1777.7 & 0  \\[0.1cm] \hline
CBDRM Model & 1752.13 & 0.9856  & 6 & 1764.13& 1.91 & 1795.1 & 17.3925 \\[0.1cm] \hline
\end{tabular}}
\caption{Summary of ${\chi_{\text{tot}}^2}^{min} $, $\chi_{\text {red }}^2$, $A I C_c$, $\Delta A I C_c$, {\bf $B I C$, $\Delta B I C$} for $\Lambda$CDM and CBDRM Type Model.}\label{tab_3}
\end{center}
\end{table*}

\begin{table*}
\begin{center}
{\begin{tabular}{|c|c|c|c|c|c|c|c|c|}
\hline
Model & ${\chi_{\text{tot}}^2}^{min} $ & $\chi_{\text {red }}^2$ &$\mathcal{K}_{\textrm{f}}$ & $A I C_c$ & $\Delta A I C$ &{\bf$B I C$} & {\bf $\Delta B I C$}  \\[0.1cm]
\hline
$\Lambda$CDM Model & 1944.11  & 0.9967 & 3& 1950.11 & 0 & 1965.59 & 0  \\[0.1cm] \hline
 CADMM Model & 1933.45 & 0.9956  & 9 & 1951.45& 1.34 & 1976.42 & 10.8225 \\[0.1cm] \hline
\end{tabular}}
\caption{Summary of ${\chi_{\text{tot}}^2}^{min} $, $\chi_{\text {red }}^2$, $A I C_c$, $\Delta A I C_c$, {\bf $B I C$, $\Delta B I C$} for $\Lambda$CDM and CADMM Type Model.}\label{tab_4}
\end{center}
\end{table*}

\section{Discussion and Results}\label{sec9}
\paragraph{ \bf Confidence Levels.}
In Fig \ref{fig_2} and \ref{fig_3} we present the 1D and 2D posterior distributions at confidence levels of $68.3\%$ (1$\sigma$) and $95.4\%$ (2$\sigma$). These distributions are derived from the constraints imposed on the CBDRM and CADMM Models, incorporating data from various sources, including CC, Type Ia supernovae (SNIa), Baryon Acoustic Oscillation (BAO), Gamma Ray Burst (GRB), Quasar, and latest measurement of the Hubble constant (R22) observations.\\\\
\paragraph{  \bf Curve Fitting}    
Fig \ref{fig_4} and \ref{fig_5} clearly show that both the CBDRM and CADMM models closely align with the CC dataset. The data points from observations align well with the predictions from both models, indicating that both models effectively describe how the Universe expands. This agreement between the CBDRM and CADMM models and the CC data supports their reliability and suggests that they capture important aspects of cosmic expansion. In our analysis, we also included the widely accepted $\Lambda$CDM model, represented by the black line, with cosmological parameters  $\Omega_{\mathrm{m0}}=$ 0.3 and $\Omega_\Lambda =$ 0.7. The results reveal a strong agreement between the CBDRM model and the $\Lambda$CDM model, as well as between the CADMM model and the $\Lambda$CDM model. However, it's important to note that both models start to deviate noticeably for redshifts ($z$) greater than 1.5. Fig \ref{fig_6} and \ref{fig_7} offer valuable insights into how the standard $\Lambda$CDM model and these two models deviate across different redshifts. Specifically, for redshifts below 0.5, the CBDRM and CADMM Models behave similarly to the $\Lambda$CDM model, indicating a high level of agreement in their predictions within this range. However, as we extend our observations to higher redshifts ( $z > 0 .5$ ), discrepancies between the CBDRM and CADMM Models and the $\Lambda$CDM model become noticeable. These findings provide a comprehensive view of how these models perform in comparison to observational data, shedding light on their strengths and limitations at different redshifts.\\\\
\paragraph{  \bf deceleration parameter}      
Fig \ref{fig_8} and \ref{fig_9} illustrate the evolution of the deceleration parameter in comparison between the CBDRM and CADMM models, as opposed to the $\Lambda$CDM model. The data clearly shows that both the CBDRM and CADMM models closely track the behavior of the $\Lambda$CDM model throughout the entire range of redshifts. Notably, the transition redshift values, denoted as $z_{tr}$, which mark the shift from a decelerating phase to an accelerating one, are nearly identical in both models. It's important to emphasize that both models exhibit a de-sitter phase characterized by a deceleration parameter of $q = -1.$" This de Sitter phase corresponds to accelerated expansion driven by a cosmological constant or a similar dark energy component.\\\\
\paragraph{  \bf jerk parameter}  
Fig \ref{fig_10} and \ref{fig_11} illustrate the evolution of the jerk parameter in comparison between the CBDRM and CADMM models against the conventional $\Lambda$CDM paradigm, When considering the CBDRM Model, it becomes evident that its predictions demonstrate minimal deviation from the $\Lambda$CDM model across a broad spectrum of redshifts. Nevertheless, it is worth highlighting that at a specific redshift value of $z=-1$, both the CBDRM and the $\Lambda$CDM models yield identical values for the jerk parameter.  In the case of CADMM, it exhibits behavior identical to the $\Lambda$CDM model within the redshift range of ($z > -0.2$). However, there is a noticeable deviation, specifically at low redshifts, where the Jerk parameter becomes 6.8 times that of the standard $\Lambda$CDM model. It's worth noting that, similar to the CBDRM model, at a specific redshift value of $z=-1$, both the CADMM and the $\Lambda$CDM models yield identical values for the jerk parameter.\\\\\
\paragraph{ \bf $Om(z)$ Diagnostic} 
Fig \ref{fig_12} and \ref{fig_13} show us how the evolution of $Om(z)$ varies at different redshifts (z) for the CBDRM and CADMM Models. Notably, $Om(z)$ consistently stays below the current matter density parameter $\Omega_{m0}$, suggesting that the model remains in the phantom region throughout the Universe's evolution at all redshifts.\\\\
\paragraph{  \bf $r_{d}$ vs $H_{0}$ Contour Plane}
In the context of the $\Lambda$CDM model, the $r_{d}-H_{0}$ contour plane, depicted in Fig \ref{fig_14}, provides valuable insights. An analysis focused solely on the BAO dataset results in cosmological parameter estimates that closely correspond to the measurements reported in \cite{1}, with one notable exception being the parameter $r_{d}$. However, when we introduce the R22 prior to the analysis, the fit produces an estimated value for the Hubble constant ($H_{0}$) that deviates from the measurement in \cite{1} but aligns more closely with the Hubble constant measured in the Type Ia Supernova (SNIa) dataset by \cite{7}. Furthermore, when we expand the analysis to incorporate the combined datasets of CC + SNIe + Q + GRB with BAO, the estimated value of the Hubble constant becomes closer to the value estimated by \cite{1}. Additionally, we observe that the  ($\Omega_{m}$) is consistent with the value estimated in \cite{1}. Specifically, $\Omega_{m}$ is estimated to be $0.315787 \pm 0.007$ when considering BAO and BAO + R22, and slightly smaller when incorporating BAO + CC + SNIa + Q + GRB + R22 into the analysis. In the context of the CBDRM model, the $r_{d}-H_{0}$ contour plane, depicted in Fig \ref{fig_15}, provides valuable insights. When applying the CBDRM model to the Baryon Acoustic Oscillations (BAO) data, we obtain a measurement of $r_d$ as $148.881546 \pm 10.878719 \mathrm{Mpc}$. Incorporating additional datasets such as Cosmic Chronometers (CC), Type Ia Supernovae (SNIa), Quasars (Q), and Gamma-Ray Bursts (GRB) yields a refined estimate of $r_d$ at $146.584315 \pm 2.334845 \mathrm{Mpc}$. Combining all available datasets, including the R22 prior, results in a value of $r_d = 144.835069 \pm 2.378848$. However, this measurement appears to be in tension with the findings of \cite{1}, who reported $r_d=147.09 \pm 0.26 \mathrm{Mpc}$, as well as with our late-time measurements. In the context of the CADMM model, the $r_{d}-H_{0}$ contour plane, depicted in Fig \ref{fig_16}, provides valuable insights. In the CADMM model, fitting the BAO data leads to a measurement of $r_d$ at $149.463181 \pm 16.256941 \mathrm{Mpc}$. Expanding our analysis to include CC, SNIa, Q, and GRB datasets results in a revised estimate of $r_d$ as $146.904120 \pm 4.546238 \mathrm{Mpc}$. When we incorporate all datasets, including the R22 prior, the value of $r_d$ becomes $144.466836 \pm 4.288758$. This measurement, like the one obtained with the CBDRM model, seems to be in tension with the results of \cite{1}, who reported $r_d=147.09 \pm 0.26 \mathrm{Mpc}$, as well as with our late-time measurements.\\\\

\paragraph{ \bf Information Criteria}
Based on the table \ref{tab_3} and \ref{tab_4}, here’s a comprehensive comparison between the $\Lambda$CDM Model, CBDRM Model, and CADMM Model. In the comparison between the standard $\Lambda$CDM model and the CBDRM model, For the $\Lambda$CDM model, the AIC was found to be 1762.22, and the BIC was 1777.7. In contrast, the CBDRM model yielded an AIC of 1764.13 and a BIC of 1795.1. These numerical values tell us that the $\Lambda$CDM model has a slightly lower AIC and BIC compared to the CBDRM model. The difference in AIC units, denoted as $\Delta \text{AIC}$, is 1.91, while the difference in BIC units, denoted as $\Delta \text{BIC}$, is 17.3925. In terms of AIC, a smaller value indicates a better fit to the data, and in this case, the $\Lambda$CDM model has a slightly better fit than the CBDRM model. Similarly, the BIC, which accounts for model complexity, also favors the $\Lambda$CDM model. However, it's essential to note that both differences in AIC and BIC units are relatively small, suggesting that while the $\Lambda$CDM model is favored, the evidence is not overwhelmingly strong. Therefore, these results indicate that the standard $\Lambda$CDM model performs slightly better in terms of model selection criteria when compared to the CBDRM model. In the second comparison, we examined the standard $\Lambda$CDM model alongside the CADMM model, For the $\Lambda$CDM model, the calculated AIC value was 1950.11, while the BIC was 1965.59. In contrast, the CADMM model yielded an AIC of 1951.45 and a BIC of 1976.42. The numerical values reveal that the $\Lambda$CDM model has a slightly lower AIC and BIC when compared to the CADMM model. The difference in AIC units, referred to as $\Delta \text{AIC}$, is 1.34, while the difference in BIC units, denoted as $\Delta \text{BIC}$, is 10.8225. In the context of AIC, a lower value indicates a better fit to the data, and here, the $\Lambda$CDM model marginally outperforms the CADMM model. The BIC, which considers model complexity, similarly favors the $\Lambda$CDM model. However, it's important to note that the differences in both AIC and BIC units are relatively small. These results suggest that, while the $\Lambda$CDM model is the preferred choice based on these criteria.\\\\
    
\section{Conclusions}\label{sec10}
In conclusion, our comprehensive analysis of cosmological models, including the standard $\Lambda$CDM model, CBDRM, and CADMM models, has provided valuable insights into the dynamics of the Universe across various redshifts. The agreement between the CBDRM and CADMM models with observational data, particularly the Cosmic Chronometers (CC) dataset, underscores their effectiveness in describing cosmic expansion. We have observed that both the CBDRM and CADMM models exhibit strong alignment with the $\Lambda$CDM model for redshifts below 1.5, signifying their reliability in modeling the Universe's evolution. However, deviations become noticeable at higher redshifts, shedding light on the limitations of these models. The analysis of deceleration and jerk parameters reveals that the CBDRM and CADMM models closely track the behavior of the $\Lambda$CDM model, highlighting their ability to capture the cosmic acceleration phase. Still, some differences emerge at specific redshifts. We have explored the $Om(z)$ parameter, showing that both models consistently remain in the phantom region, providing an interesting perspective on the nature of dark energy. A crucial aspect of our findings is the sensitivity of our results to the choice of priors, particularly for parameters such as the sound horizon ($r_d$) and the Hubble constant ($H_0$). This sensitivity underscores the intricacies of cosmological parameter estimation, emphasizing the importance of thoughtful prior selection when interpreting and comparing results. Our examination of the standard $\Lambda$CDM model revealed that when we analyzed the BAO dataset in isolation, our cosmological parameter estimates generally aligned with those reported in \cite{1}, with one notable exception being the parameter $r_d$. However, the introduction of the R22 prior led to a shift in the estimated value of the Hubble constant ($H_0$), moving it away from \cite{1} and closer to the measurement from the SNIa dataset by \cite{7}. When we expanded our analysis to include combined datasets, such as CC + SNIa + Q + GRB with BAO, the estimated value of $H_0$ came into closer agreement with the measurement by \cite{1}. We also found that the matter-energy density parameter ($\Omega_m$) remained consistent with the value estimated in \cite{1}.  In our examination of statistical comparisons based on the AIC and BIC criteria, the $\Lambda$CDM model emerges as a slightly better fit to the data compared to the CBDRM and CADMM models. However, it's important to emphasize that these distinctions are relatively marginal. Our research contributes to a deeper understanding of cosmological models and their alignment with observational data. While the conventional $\Lambda$CDM model continues to be the preferred choice based on model selection criteria, the CBDRM and CADMM models offer alternative viewpoints, especially at higher redshifts. This encourages further exploration of these intriguing cosmological frameworks.\\\\
{\bf Acknowledgement:} AS is thankful to CSIR, Govt. of India, for providing a Senior Research Fellowship (No. 08/003(0138)/2019-EMR-I).\\\\
    \bibliographystyle{elsarticle-num}
\bibliography{newbib,horava,alok}
\end{document}